\documentclass[preprint]{aastex}
\begin{document}
\received{}
\accepted{}
\revised{}
\title{The Changing Blazhko Effect of XZ Cygni}
\author{Aaron LaCluyz\'{e}, Horace A. Smith,  E.-M. Gill, A. Hedden 
\altaffilmark{1},
Karen Kinemuchi, A. M. Rosas\altaffilmark{2}, Barton J. Pritzl\altaffilmark{3},
 Brian Sharpee
\altaffilmark{4}, 
Christopher Wilkinson\altaffilmark{5}, and K.W. Robinson}
\affil{Department of Physics and Astronomy, Michigan State University,
East Lansing, MI 48824}
\author{Marvin E. Baldwin and Gerard Samolyk}
\affil{AAVSO RR Lyrae Committee, 25 Birch St., Cambridge, MA 02138}

\altaffiltext{1}{Current address: Dept. of Astronomy, Steward Observatory,
Tucson, AZ 85721}
\altaffiltext{2}{Current address: Catholic University of America,
NASA/GSFC, Mail Code 695.0, Greenbelt, MD 20771}
\altaffiltext{3}{Current address: Dept. of Physics and Astronomy, 
Macalester College, Saint Paul, MN 55105}
\altaffiltext{4}{Current address: Molecular Physics Laboratory, PN-099,
SRI International, Menlo Park, CA 94025}
\altaffiltext{5}{McDonald Observatory, University of Texas at Austin,
1 University Station, Austin, TX 78731}

\begin{abstract}

New CCD photometry has been obtained for the RR Lyrae variable star
XZ Cygni.  An analysis of old and new photometry confirms
earlier results that XZ Cygni exhibits the Blazhko Effect and that its
Blazhko period has changed over time.  These changes
in the Blazhko period are anticorrelated with observed changes in the
primary period of XZ Cygni.  During the first half of the twentieth
century, XZ Cygni had a Blazhko period of approximately 57.4~d. 
Beginning in 1965, its primary period underwent a steep
decline in several steps.  Coincidentally, its Blazhko period
increased to about 58.5~d.  In 1979, the primary period
suddenly increased again.  After an interval in which
the Blazhko Effect was small, the Blazhko Effect re-established
itself, with a period of approximately 57.5~d. When its Blazhko period
is near 57.5~d, XZ Cygni has also shown a tertiary period of 41.6~d.
We confirm that there is evidence for a longer 3540~d
period in photometry obtained during the first
half of the twentieth century.  XZ Cygni is compared with three
other RR Lyrae stars which also appear to show changing Blazhko periods.
The observed changes in the length of the Blazhko period of XZ Cygni
constrain possible explanations for the Blazhko Effect. In
particular, they argue against any
theoretical explanation which requires that
the Blazhko period be exactly equal to or directly proportional to
the rotation period of the star.

\end{abstract}


\newpage

\section{Introduction}

\citet{Bl07} discovered
a periodic oscillation in the times
of maximum light of the RR Lyrae variable RW Dra.  \citet{Sha16} found 
that the light curve shape of
RR Lyrae itself changed with a secondary period of approximately 40
days.  Subsequently, long term secondary modulations in
light curves have been identified for a number of well-observed
RR Lyrae variables. These secondary modulations are now usually
referred to as the Blazhko Effect, although
\citet{Sze00} have suggested that the name Blazhko-Shapley Effect
might be appropriate in recognition of Shapley's role in their discovery.
In the Milky Way, about 20 - 30\% of
RR Lyrae variables
of Bailey type ab (RRab, also termed RR0) show the Blazhko Effect
\citep{M03,Sz88}.  In the Large Magellanic Cloud the percentage of
RRab stars which show the Blazhko Effect is smaller, about 12\% to 15\% 
\citep{A03,So03}.
Periods of 
reported secondary modulations of classical Blazhko Effect RRab stars
in the Milky Way
range from 11~d to 533~d. More recently, secondary
modulations similar or identical to the Blazhko Effect have
also been discovered for a small percentage of the RR Lyrae
stars of Bailey type c (RRc or RR1). Reviews of the Blazhko Effect have been
given by \citet{Sz76}, \citet{Sz88}, \citet{Sm95} and \citet{Kov01}. 
As we will discuss in $\S$7,
several explanations have been advanced, but there is still no consensus 
on the mechanism underlying the Blazhko phenomenon \citep{co83,Shi00,Kov01}. 

Many RR Lyrae stars are known to have undergone changes in the length
of their primary pulsation cycle \citep{Sm95}.  However, significant changes in
the length of the Blazhko period appear to have been reported for only four
stars \citep{J02}: XZ Dra, RV UMa, RW Dra,
and the subject of this study, XZ Cyg (HD 239124, BD +56 2257).  
RR Lyrae stars with changing Blazhko periods are of particular
interest because they may shed light on the cause of the Blazhko
Effect.

The variability of XZ Cygni was discovered in 1905
by L. Ceraski.  Subsequent investigations have made XZ Cyg one of the best 
studied
RRab stars.  \citet{Bl22} established that XZ Cygni displays the
Blazhko Effect, finding the Blazhko period to be approximately
57.4~d long. For the next half century, investigations of XZ Cyg confirmed the
existence of a Blazhko period near 57.4~d, though some indicated
the possibility of a 41.6~d tertiary period as well \citep{K58,M53}.
In the 1960s, however, XZ Cyg underwent striking changes.
\citet{B73}, \citet{S75}, \citet{P75}, and \citet{MT75}
reported that the Blazhko period
of XZ Cyg abruptly lengthened to about 58.4~d in the mid 1960s.
This change in the Blazhko period was coincident with a
sharp decrease in the primary pulsation period, which took place
in several steps \citep{Be88,Ba03}.  More
recently, the primary period has increased once more,
rising to a value
close to the one it had before its sudden decline \citep{Ba03}.

In the present paper we investigate the behavior of XZ Cyg between its
discovery and 2002.  We will present new CCD photometry of XZ Cyg
and analyze old and new observations of the star. Our intent is
to verify that the Blazhko period of XZ Cyg has indeed changed and
to understand how any changes in the Blazhko period are related to
changes in the primary period.  Finally, we will consider how
observed changes in the Blazhko period might constrain
theoretical models of the Blazhko phenomenon.

\section{CCD Observations}

Differential $V$ band photometry of XZ Cyg was obtained with the 60-cm
telescope of the Michigan State University Observatory between 
JD 2451338 and 2452506 (1999 June 8 through 2002 August 19).  
Until JD 2452202 observations
were obtained with an Apogee Ap7
CCD camera.  After that date, an Apogee Ap47p camera was used instead.
The star TYC 3929-1703-1 was adopted as a primary comparison star
and TYC 3929-1811-1 was used as a check star.  Transforming the Tycho catalog
$VT$ and $BT$ magnitudes to the Johnson system using the prescription
of \citet{Be00}, one obtains $V$ = 9.97 $\pm$ 0.04 and $\bv$ = 1.31 $\pm$ 0.08
for TYC 3929-1703-1. Differential V-band
photometry for XZ Cyg is listed in Table 1. Differential
magnitudes are given in the sense $V$(XZ Cyg) minus $V$(comparison
star). The typical uncertainty in a
measurment of $\Delta V$ is 0.01 mag. A total of 7738 observations of
$\Delta$V were obtained.

The comparison star is redder than XZ Cyg
itself, for which $\bv$ varies between 0.2 and 0.4 \citep{S66}.  Because we are
interested mainly in analyzing our observations to determine the length
of the Blazhko cycle, and because we do not have multi-filter
observations, we have not attempted to apply any color corrections 
to observed
values of $\Delta V$.  From the previously observed color
range of XZ Cyg, we would expect such corrections to make 
the observed values
of $\Delta V$ smaller by about 0.03 mag near maximum light and by 0.02 mag
near minimum light for both the Ap7 and Ap47p observations. Corrections
of that size are not important for our purposes.

441 additional measurements of $\Delta$V were obtained between 
JD 2452210 and 2452601 (2001 October 27 to 2002 November 22)
by G. Samolyk, who used a 25 cm telescope of the Milwaukee Astronomical
Society Observatory with a Santa Barbara Instrument Group ST9E CCD.
The uncertainty of a typical measurement of $\Delta$V with this instrument
is about $\pm$0.02 mag.  Simultaneous observations of XZ Cyg
indicate that the $V$ band photometric systems of the Michigan State
University observations and the Milwaukee observations are the same to
within $\pm$0.01 mag. The Milwaukee differential $V$ measurements are listed
in Table 2. The same comparison star was used as for the Michigan State
University observations.

\section{Analysis of the CCD Observations}

The period search routines CLEANest \citep{F95}, Period98 \citep{S98}, and PDM
\citep{St78} were used to determine the primary period of XZ Cyg based
upon the full CCD dataset.  The three methods gave consistent
results, $P_0$ = 0.466598 $\pm$ 0.000002~d,
a value also consistent with that recently determined by \citet{Ba03} from AAVSO
observations spanning the interval HJD 2448570 to 2452618: 0.46659934
$\pm$ 0.00000035~d.
The light curve of XZ Cyg, folded according to the \citet{Ba03} ephemeris, is shown in
Figure 1 (for the Michigan State University observations only) and Figure
2 (for all of the CCD observations). These figures strongly indicate
the existence of a continued Blazhko Effect in XZ Cyg during the time
interval in which the CCD data were obtained. 

We then searched for possible additional periods in the CCD dataset
using the CLEANest and Period98 programs.  
The data were prewhitened to remove the
primary frequency, $f_0$, and its nine higher harmonics (2$f_0$,
3$f_0$,...10$f_0$).  Prewhitening the data in that way left
residuals with a standard deviation of 0.079 mag. The Blazhko Effect
reveals itself in two different types of frequency structure.  Some
Blazhko variables show frequency triplets, with peaks at frequencies 
$jf_0 \pm kf_B$, where j = 1, 2, 3,..., k = 0 or 1, and $f_B$ is the frequency
of the Blazhko period.  Other Blazhko variables show only frequency
doublets (see, for example, \citet{A03}). Our observations of XZ Cygni 
indicate the presence of
frequency triplets, although the presence
of frequency aliases complicates the Fourier spectrum.
A portion of the Fourier spectrum around the frequency of the primary
period is shown in Figure 3 for the prewhitened data.  The peaks at 
frequencies 2.12577 c/d
and 2.16054 c/d would beat with the primary frequency at periods
of 57.45~d and 57.59~d, respectively.  Similarly spaced peaks are
evident in the Fourier spectrum near frequency $2f_0$. Overall, the
peaks identified in the CLEANest and Period98 power spectra indicate
the presence of a Blazhko period of 57.5 $\pm$ 0.2~d.
Subtraction of frequencies $jf_0 \pm f_B$, where j = 1,2,..10 and
$f_B$ = 0.01739 reduced the residuals to 0.042 mag, still significantly
larger than the expected observational uncertainty.

With the data prewhitened to remove the primary and Blazhko periods, 
a further search was conducted for peaks in the Fourier spectrum
which might represent a tertiary period.  A portion of the resultant
Fourier spectrum is shown in Figure 4. The highest peaks occur at
frequencies of 2.1191 c/d and 2.1698 c/d, which would beat with 
the primary frequency
at intervals of 41.6~d and 37.5~d, respectively. A lesser peak
is evident at a frequency of 2.1671 c/d, which would beat with the
primary frequency at an interval of 41.8 c/d. Note that the
frequency at 2.1698 c/d can be interpreted as a one year alias
of the frequency at 2.1671 c/d. The Fourier spectrum
around the location of $2f_0$ shows peaks at frequencies of 4.2623 c/d
and 4.3104 c/d,
which would be indicative of a beat period of length 41.6~d. A peak is also
evident at a frequency of 4.2596 c/d, which is, however, interpetable as
a one year alias of the peak at 4.2623 c/d. We thus conclude that the CCD
photometry indicates the existence of a tertiary period of length 41.6~d
with an uncertainty of about $\pm$ 0.2~d.
As will be discussed later, this period is
close to the tertiary periods found by \citet{M53} and \citet{K58}.  Period98
was used to fit the data with a set of frequencies of the form
$jf_0, jf_0 \pm f_B$,  and $jf_0 \pm f_3$, where $f_3$ is the frequency of 
a 41.6~d
tertiary period, 0.02404 c/d. The resultant fit left residuals of 0.029
mag, still somewhat larger than expected from observational error
alone.  A search for additional periodicities did not lead to clear
evidence that any exist.  A portion of the Fourier spectrum
around the primary frequency after removal of a 0.466598~d primary period, 
the 57.5~d
secondary period, and
41.6~d tertiary period is shown in Figure 5. The frequencies, amplitudes, and phases
of the components of this fit are listed in Table 3.

The period search procedure was repeated incorporating just the
Michigan State University Observations.
The results, as might be expected, are very similar. The residuals in the
final fit to the Michigan State University observations
were slightly smaller, $\pm$ 0.026 mag.

The effects of the secondary and tertiary periods have been removed
from the lightcurve by subtracting the sidelobe components of the 57.5~d
and 41.6~d periodicities.  The resultant light curves, phased as in
Figures 1 and 2, are shown in Figures 6 and 7.

\section{XZ Cygni before 1965}

The most extensive studies of the behavior of XZ Cyg before the changes
in primary period which began about 1965 are those of \citet{M53} and
\citet{K58}.
\citet{M53} obtained the first detailed photoelectric observations of XZ
Cyg, monitoring the star from 1948 until November 1952.  \citet{M53}
confirmed the existence of the secondary period reported by \citet{Bl22},
finding the period to be approximately 57.4~d.
However, \citet{M53} found that the observations were better described if
in addition to this secondary period there existed a tertiary period
with a length of either 89.34 or 94.31 times the primary period, 41.7~d
or 44.0~d, respectively, assuming a primary period of 0.4665839~d.

\citet{K58} attempted, in so far as was possible, to put all existing
observations of the maximum brightness of XZ Cyg on a consistent photometric
system.  \citet{K58} then used the times and brightnesses of maxima obtained
between JD 2417017 and JD 2434946 to study the long term behavior
of XZ Cyg.  \citet{K58} found evidence for a small decline in the primary
period of the variable during this time span, described by the
ephemeris for the times of maximum light

\begin{eqnarray}
Max (HJD) = 2417201.241 + 0.4665878E - 0.000107E^2\times 10^{-6} \
\end{eqnarray}

\noindent where E is the number of cycles elapsed since the start date.
Thus, the period would have decreased from 0.4665878~d in 1905 to 
0.466584~d in 1954.  \citet{Be88} provided a slightly
different interpretation of the period changes of XZ Cyg during
this interval, finding that the primary
period decreased in a stepwise fashion from 0.4665861~d during
JD 2417000 - 2424800 to 0.466579~d during the interval JD 2434600 -
2438700 (ending in 1964). Whichever description one accepts for the period 
changes, 
it is clear that the primary pulsation 
period of XZ Cyg was relatively stable between 1905 and 1965.

\citet{K58}
confirmed the existence of a Blazhko period of 123.040 times
the primary period (giving $P_{Bl}$ = 57.4~d).  \citet{K58} also confirmed
the existence of a tertiary period of 89.23 times the primary period
($P_{3}$ = 41.6~d).  Thus, three times the secondary period would be
approximately (but not exactly) four times the tertiary period, as \citet{K58} 
noticed. \citet{K58} also identified a superposed, longer quaternary variation 
with
a period of about 3460~d, during which there is a long term change in
the mean magnitude at maximum light.  This can be seen in Figure 8,
which plots the magnitude at maximum light versus Julian Date for the
maxima in Table 10 of \citet{K58}. Not plotted in Figure 8 are the later
maxima listed by \citet{K65}, which are approximately on the system 
of \citet{K58}
but which may not be exactly comparable.
Long term periodicities have been reported for a few other Blazhko Effect
stars, most notably RR Lyrae itself \citep{Det73}.  In addition to its
41~d Blazhko period, RR Lyrae exhibits a four year periodicity.  The
amplitude of the Blazhko Effect varies during this longer cycle and the
phase of the Blazhko cycle shifts when one four-year cycle ends and the next begins.
The data accumulated by \citet{K58} indicate that the amplitude of the
Blazhko Effect in XZ Cyg changed over time, but they are not sufficient to
let us decide whether the 3460~d period is exactly analogous to RR Lyrae's
four year period. \citet{D56} reported that J. Balazs had
identified a 153.8~d tertiary
period for XZ Cyg, which, however, might also be interpreted as the
beat interval between periods of approximately 57.4~d and 41.6~d.

As a check on the conclusions of \citet{K58} and \citet{M53}, the 
CLEANest routine
was used to search for periodicities in the magnitudes of maximum
reported by \citet{K58}. Applied to the entire data range from JD 2417017 to
JD 2434946, the CLEANest routine did identify periods of 57.41 ($\pm$0.02)~d,
41.61($\pm$0.04)~d, and 3540($\pm$100)~d.  Thus, we broadly confirm
\citet{K58}'s results. Nonetheless, it is also evident that these three
periodicities are not able to explain much of the scatter in the
data.  The standard deviation of the maximum
magnitudes about the mean was initially 0.113 mag.  Removing these three periods 
and their 
harmonics from the data reduced that value only slightly, to 0.086 mag.
It is
possible that instead of being 3540~d long,  the actual longest
period could be twice that value.  When the data are phased with a circa 
7000~d period, there
are large gaps in the phase coverage which make it difficult to fully
test that possibility.
In Figure 9, we plot the residuals of maximum magnitudes from \citet{K58} after
removal of the three identified periodicities.  Moreover, when
subsamples of the data are examined, it is clear that the strength of
the secondary period has not been constant over the entire duration
of the observations collected by \citet{K58}.  There is, for example,
little evidence for a circa 57~d period in the observations made from
JD 2417017 to 2417538.  A period near 57.34~d is, however, present 
in the
observations made between JD 2424370 and JD 2427750 (Figure 10). Such 
variability
in the Blazhko Effect would, of course, not be unique to XZ Cyg, being also
well observed in RR Lyrae itself, for example. In
addition, despite the efforts by \citet{K58} to place all observations
on a consistent system, the heterogeneity of the observations makes
it difficult to compare results over the entire time interval
for which \citet{K58} tabulated maxima.

For the interval spanned by the observations of \citet{M53}
and the data compiled by \citet{K58}, one can conclude that the primary 
period of XZ Cyg was decreasing slightly but was relatively stable, and
that additional periods of length about 41.6~d, 57.4~d, and 3540~d were
present. The durability of the 3540~d period is
uncertain, since it could be followed for only a few cycles, and, as we have
noted, the possibility that the longest period is 7000~d rather than
3540~d cannot be entirely excluded.
Moreover, it appears likely that the amplitudes of the 41.6~d and 57.4~d 
periodicities were not constant throughout that time interval.  
Nonetheless, our check confirms prior studies of the
behavior of XZ Cyg before 1965.

\section{The behavior of XZ Cygni since 1965}

Beginning about JD 2438800 (1965), the primary period of XZ Cyg began a 
steep decline
in several steps \citep{Be88,Ba03}. The primary period fell to a minimum
of 0.4664464~d during the interval JD 2442050 to
JD 2443740, 1974 to 1978 \citep{Ba03}. In 1979, the primary period sharply 
increased,
rising to values slightly greater than those observed before the steep decline.
Changes in the primary period of XZ Cyg since 1965 are listed in
Table 4, which follows \citet{Ba03}. The periods reported in \citet{Ba03}
are similar to those reported by \citet{Be88} for time intervals
covered in both studies. 

Shortly after the decline in the primary period of XZ Cyg, observers
began to report a coincident increase in its 57.4~d Blazhko period
\citep{B73,P75,K75,S75}.  These indicated that, at
about the time that the primary period of XZ Cyg fell to its minimum, the
Blazhko period of XZ Cyg increased to about 58.4~d, or perhaps slightly
more. An O-C diagram illustrating this change in the Blazhko period
is included in \citet{S75}. As a check and an extension of these earlier results, we have
analyzed observations of XZ Cyg reported since 1965 to the RR Lyrae
Committee of the American Association of Variable Star Observers (AAVSO).

\subsection{AAVSO visual observations}

AAVSO observers have accumulated a large number of visual observations of
XZ Cyg since 1965. These data mainly consist of observations obtained
near the time of maximum light, rather than equally covering the entire
light cycle. We begin by considering M. Baldwin's observations of the
magnitude attained by XZ Cyg at maximum light \citep{Ba02}.
In Figure 11, we plot the magnitude of maximum light against Julian Date.  
It is apparent that
the range of scatter in the maxima was relatively large around JD 2443000,
but diminished after 2444000.  This change is also evident in
the plots of O-C (observed minus
calculated times of maximum light) in \citet{Ba03}, which are corrected for
the observed changes in the primary period of the star. 
The diminution in the scatter was noticed by \citet{T80} who
further noted that it was approximately coincident with the abrupt large
increase in the primary period of XZ Cyg which took place near JD 2444000. 
Subsequent to
JD 2449000, there is some evidence for an increase again in the size of the
scatter of
the maximum magnitudes, but not to the level observed near JD 2443000.

We have made two new analyses of the AAVSO observations to determine
the length and amplitude of the Blazhko cycle during the time interval covered
by the AAVSO observations.
We first searched for a periodicity in the magnitude reached at maximum
light using the CLEANest routine, limiting ourselves to the maxima
observed by M. Baldwin to reduce the possibility of systematic
errors among different visual observers.  Results from this analysis are 
listed in
Table 5. We also searched for periodicities in the O-C values provided by
\citet{Ba03}.  In this case, we analyzed all of the O-C values reported
by AAVSO observers. Results from a period search using the CLEANest program
are listed in Table 6. For comparison, in Table 7 we list a number
of determinations of the Blazhko period of XZ Cyg from the literature.

These analyses confirm that in the interval from about JD 2439000 to
2444000 the Blazhko period was indeed about a day longer than it had
been in the time intervals studied by \citet{M53} and \citet{K58}. The
Blazhko amplitude was particularly large around JD 2443000, when the
primary period of XZ Cyg was near its minimum.   The Blazhko periodicity was
apparently suppressed for awhile when the primary period abruptly
increased near JD 2444050, before gradually starting to reassert 
itself. The renewed
Blazhko period is shorter again. However, the evidence for the
current length of the Blazhko period is stronger in the CCD observations
than in the AAVSO visual data.

It would be very interesting to know whether the 41.6~d tertiary
period was still present at the time that the secondary period was near
58.5~d. To investigate this we analyzed AAVSO data obtained between
JD 2438882 and JD 2443800.  Figure 12 shows the Fourier spectrum for
the 109 maximum magnitudes observed by Baldwin during this time interval.
A strong peak is evident at a frequency of 0.01709 c/d, corresponding to a
period of 58.5~d.  Figure 13 plots the maximum magnitudes, phased
with the 58.5~d period.  For comparison, the same maxima phased with a
57.4~d period are plotted in Figure 14.  Figure 15 shows the Fourier spectrum of
the data after the magnitudes at maximum have been prewhitened to
remove the 58.5~d periodicity and its four highest harmonics.  No clear peak
is evident near a frequency of 0.024 c/d, corresponding to
a periodicity of 41.6~d.  The two highest peaks occur at frequencies
of 0.019377 c/d and 0.016345 c/d, corresponding to periods of 
51.6~d and 61.2~d, respectively. These periods would beat against a
58.5~d period in about 440 days and 1300 days, respectively.  They
are, however, much weaker than the 58.5~d Blazhko period, and it is
doubtful whether they correspond to physically real periodicities.

A similar analysis was performed on 182 O-C values observed during the
same interval, excluding two values which were very deviant from the
trend of the other observations.  The Fourier spectrum for these
observations (Figure 16) shows a peak at 0.01706 c/d, corresponding
to a period of 58.6~d, which is consistent within the uncertainties
with the value seen in Figure 12. The O-C values are phased with a
58.5~d period in Figure 17, which shows considerably more scatter than
in the corresponding diagram for magnitude at maximum light.  For
comparison, Figure 18 shows the O-C values phased with a 57.4~d
period. Prewhitening
to remove the 0.01706 c/d frequency again left no clearly significant
peaks in the Fourier spectrum.
Thus, while we find a Blazhko period of approximately 58.5~d to be present
in both the times and brightnesses of maximum light during this interval,
we find no evidence for a tertiary period. Any tertiary periodicity
present at this time must have had an amplitude less than about a third
that of the 58.5~d Blazhko cycle. In contrast, Figure 19 shows the
Fourier spectrum for O-C observations obtained between JD 2448570 and
JD 2452618.  The peaks at frequencies of 0.01736 c/d and 0.02404 c/d indicate
the existence of 57.6~d and 41.6~d periods.  A peak is also
evident at frequency 0.0146 c/d, corresponding to a period of 68.5~d.
That, however, is the one year alias of the 57.6~d period.

Figure 20 plots the Blazhko period versus the fundamental mode period
for XZ Cyg.  Here we have used our results for the length of the
Blazhko period and have adopted the results of \citet{Ba03} for the
length of the primary period.  For the earlier maxima listed by \citet{K58},
we have adopted an average primary period of 0.466584~d. Note that,
while the Blazhko period clearly is correlated with the fundamental
mode period, the exact form of that correlation is not clear.  There
might be a linear correlation, with $dP_{Bl}/dP_0$ = -8 $\pm 1 \times 10^{3}$.
On the other hand, the Blazhko period might have jumped discretely
from a value near 57.4~d when $P_0$ was near 0.4666~d to a value
near 58.5~d when $P_0$ was smaller than 0.46655~d.

\section{Comparison with other RR Lyrae stars having changing
Blazhko periods}

As noted in \S 1, changes in the Blazhko period have
been reported for three RR Lyrae stars besides XZ Cyg \citep{J02}.
In all four reported cases of changing Blazhko periods, the change in
the Blazhko period is coincident to changes in the primary pulsation
period.  However, the ratios of the changes in the Blazhko period to
those of the primary period, as well as the sign of that ratio,
differ from star to star. In the case of XZ Dra \citep{J02}, the
primary period and the Blazhko period change with the same sign.  
In the other three cases, RV UMa \citep{K76}, RW Dra \citep{F78},
and XZ Cyg,
the reported changes of the Blazhko period have a sign opposite
that of the changes in the primary period.  Of the four stars,
XZ Cygni has undergone the largest changes in its primary period.
Observed ratios of $dP_{Bl}/dP_{0}$ are listed in Table 8.
For the purposes of this table, we have assumed a linear dependency
of the change in the Blazhko period upon the change of the primary
(fundamental mode) period.  Such a linear dependency is consistent
with the observations, but, as we have noted above, other 
dependencies are also possible.
In the case of RW Dra, the changes in period have been estimated
from the O-C diagram of \citet{F78}. As is the case for XZ Cyg,
very long additional periods have been reported for two of the
remaining three stars. \citet{J02} identified a 7200~d 
tertiary period in XZ Dra, while \citet{K76} reported that RV UMa also shows 
a long term cycle of 2,000 to 3,000 days.

\section{Constraints on the nature of the Blazhko Phenomenon}

As also noted in \S 1, there is of yet no completely satisfactory
explanation for the Blazhko phenomenon. Proposed explanations
can be broadly divided into two groups: (1) magnetic models, in
which Blazhko variables have a magnetic field with an axis tilted
obliquely with respect to their axis of rotation, and (2) 
resonance models, in which there is a nonlinear resonance between
the dominant radial mode and a nonradial pulsation mode. Oblique
magnetic rotator models \citep{co83, Shi00} generally require that
the Blazhko period and the rotation period of the star are equal.
\citet{B58} and \citet{R94} reported the presence of a magnetic
field in the photosphere of the brightest Blazhko variable,
RR Lyrae itself.  However, \citet{P67} and 
\citet{Ch03} have not detected any strong magnetic field
in RR Lyrae.
The recent results by \citet{Ch03} have not yet been published
in full detail, but if the preliminary results are confirmed
they would provide perhaps the strongest
argument against
any oblique magnetic rotator model. Oblique rotator models have
also predicted the existence of a quintuplet rather than a triplet
structure in frequency \citep{Shi00}.  We do not see such a
structure in XZ Cygni, nor has it been identified in largescale
surveys of Blazhko variables \citep{So03, M03, A03}.

\citet{N01} and \citet{N02} have modeled the excitation of nonradial
pulsation modes in RR Lyrae stars.  They suggested that excitation
of an {\it l} = 1, {\it m} =$\pm$1 pair might result in enough amplitude and
phase modulation in the light curve to produce the Blazhko Effect.
In their models, rotational splitting of the {\it m} = $\pm$1 modes
produces the sidelobes of frequency triplets in the Fourier
spectra of photometric observations.  The Blazhko period in this model is
fixed by the rotation rate of the star, weighted with the
Brunt-V\"ais\"al\"a frequency \citep{Ha94} in the deepest part
of the radiative envelope.  Thus, there would
still exist in this model a dependency of the Blazhko period
upon the rotation period of the star.

It is difficult to test models of nonradial mode resonance on the
basis of photometric data alone, although \citet{Ko2} and
\citet{Ko3} have shown that a combination of photometry and 
high resolution spectroscopic data can constrain possible
nonradial modes.  However, the observed changes in the Blazhko period
of XZ Cyg and other RR Lyrae variables allow additional constraints
to be imposed upon theoretical explanations for the Blazhko
phenomenon.

Changes in the Blazhko period have been observed for at least four
RR Lyrae variables, as noted above.  In the case of XZ Cyg, the
length of the Blazhko period increased by about 2\% before returning
to something close to its original value. An increase and
later decrease in the rotation period of XZ Cyg by as much as
2\% is physically implausible. Moreover, while a large increase in the radius
of XZ Cyg around 1965 might produce a longer rotation period, 
the pulsation
equation, $P\sqrt{\rho} = Q$, would then indicate that the primary period
should have increased, not decreased, at that time. We would therefore argue
that the Blazhko period of XZ Cyg is unlikely to be equal to or
directly proportional with its period of rotation.  That is another
argument against simple oblique magnetic rotator models.  It may also
be an argument against the nonradial excitation models of \citet{N01}
and \citet{N02}, although there the argument is not so clear. If the
rotationally split {\it m} =$\pm$1 modes could manage to maintain a relatively
constant frequency at the time that the frequency of the fundamental
radial mode changed, then it is possible for the beat period of the
radial and nonradial modes to change so as to produce the observed
changes in the Blazhko period.

In the case of XZ Cyg, there are additional observations that 
any complete theory
must explain.  First is the existence of a 41.6~d tertiary period.
This period exists at the same time as the main 57.5~d
Blazhko period.  It might be explained through the excitation of
an additional nonradial mode.  That mode may not have been excited
during the interval when the Blazhko period of XZ Cyg was near 58.5~d.
Second, we have the coincidence in the changes of the Blazhko period
and of the period of the fundamental radial mode.  Third,
the Blazhko Effect in XZ Cyg continued strongly after
the decline in the fundamental mode period of the star, but seems
to have diminished when the fundamental period increased again.
One might imagine that the Blazhko Effect was somehow turned off
when the primary period increased, taking some years to become
re-established.  The possibility that this simultaneous change in the Blazhko
amplitude and the primary period was coincidental seems unlikely, but
cannot be entirely excluded. The Blazhko
Effect in XZ Cyg and other RR Lyrae stars has undergone changes in
amplitude even when the primary period of the star has been relatively
stable.  How the multiyear periods seen in XZ Cyg and
other RR Lyrae stars are related to the Blazhko phenomenon also remains
unclear.

It is interesting that XZ Cyg seems to have returned to
something close to the state it had during the first half of
the twentieth century following several decades of large period
changes.  It has long been known that evolution cannot account
for all of the observed changes in the primary periods of RR
Lyrae variable stars (see the review in \citet{Sm95}).  Some sort
of period change noise appears to overlie any evolutionary
period changes in these stars.  It has been suggested that
this noise might be related to discrete mixing events associated
with the semiconvective zone of the stellar core \citep{Sw79}.
Whatever is responsible for the period change noise, in the case
of XZ Cyg that mechanism was capable of reversing its behavior 
in a way that seems to have
returned the structure of the star to something close to the state
it had half a century before.

\section{Conclusions}

We confirm that the Blazhko period of XZ Cyg increased
sharply from about 57.4~d to 58.5~d at the same time
that its primary pulsation period underwent a sharp decline.
Subsequently, when the primary pulsation period increased
once more, the Blazhko period decreased.  During the
intervals when XZ Cyg has had a Blazhko period near 57.4~d,
it has also shown a tertiary period with a length of 41.6~d.
When the Blazhko period was near 58.5~d,
the 41.6~d tertiary period either was not present or had an
amplitude much smaller than that of the 58.5~d periodicity.
The amplitude as well as the period of the Blazhko Effect
in XZ Cyg has changed over time.  We particularly note the
disappearance or diminution of the Blazhko Effect around JD
2444000, at about the same time that the primary period
jumped from 0.4664464~d to 0.4666938~d.  The Blazhko Effect
is evident again after about JD 2448570, at which time
the primary period had decreased to 0.466599~d. During the first
half of the twentieth century, photometry of XZ Cyg also showed
evidence for a fourth period, in this case almost 10 years
in length.
The observed changes in the Blazhko period of XZ Cyg and a few other
Blazhko stars argue that the
Blazhko period is not exactly equal to or directly
proportional to the rotation period of the star, as required
by some theoretical models.

\begin{acknowledgements}

This work has been supported in part by the National Science Foundation
under grants AST-9986943 and AST-0205813.  We thank the American
Association of Variable Star Observers for providing a downloadable
version of the CLEANest program on their web site.  HAS thanks
the Osservatorio Astronomico di Roma and Giuseppe Bono for hospitality during a
sabbatical leave. E.M. Gill and A. Hedden were supported by the
National Science Foundation Research Experience for Undergraduates
program, under grant PHY-9912212. We thank Katrien Kolenberg and Johanna
Jurcsik for helpful comments and suggestions.

\end{acknowledgements}

\clearpage
\begin{figure}
\plotone{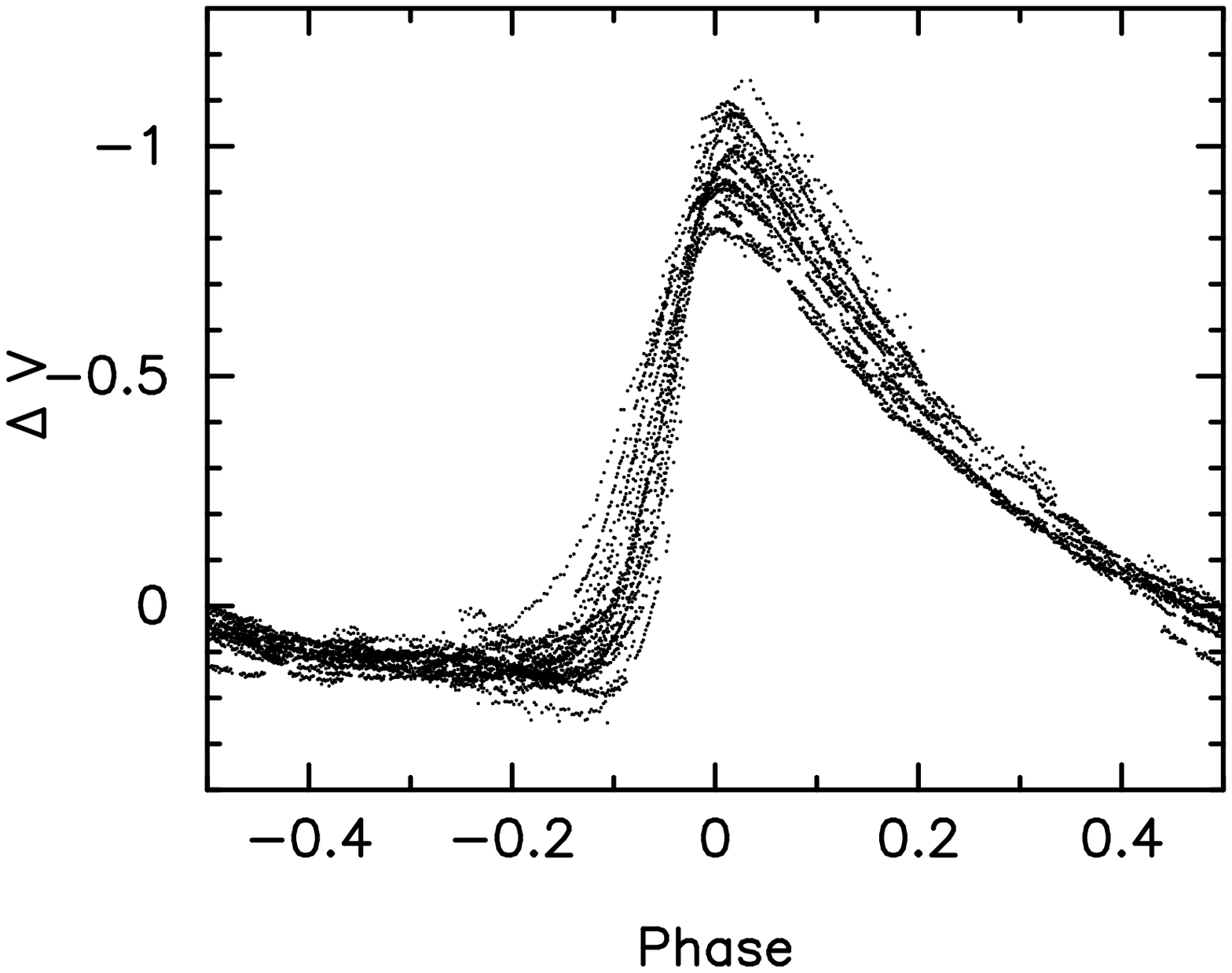}
\caption{$V$ light curve of XZ Cygni based upon
observations obtained at Michigan State University.}  
\end{figure}

\clearpage 
\begin{figure}
\plotone{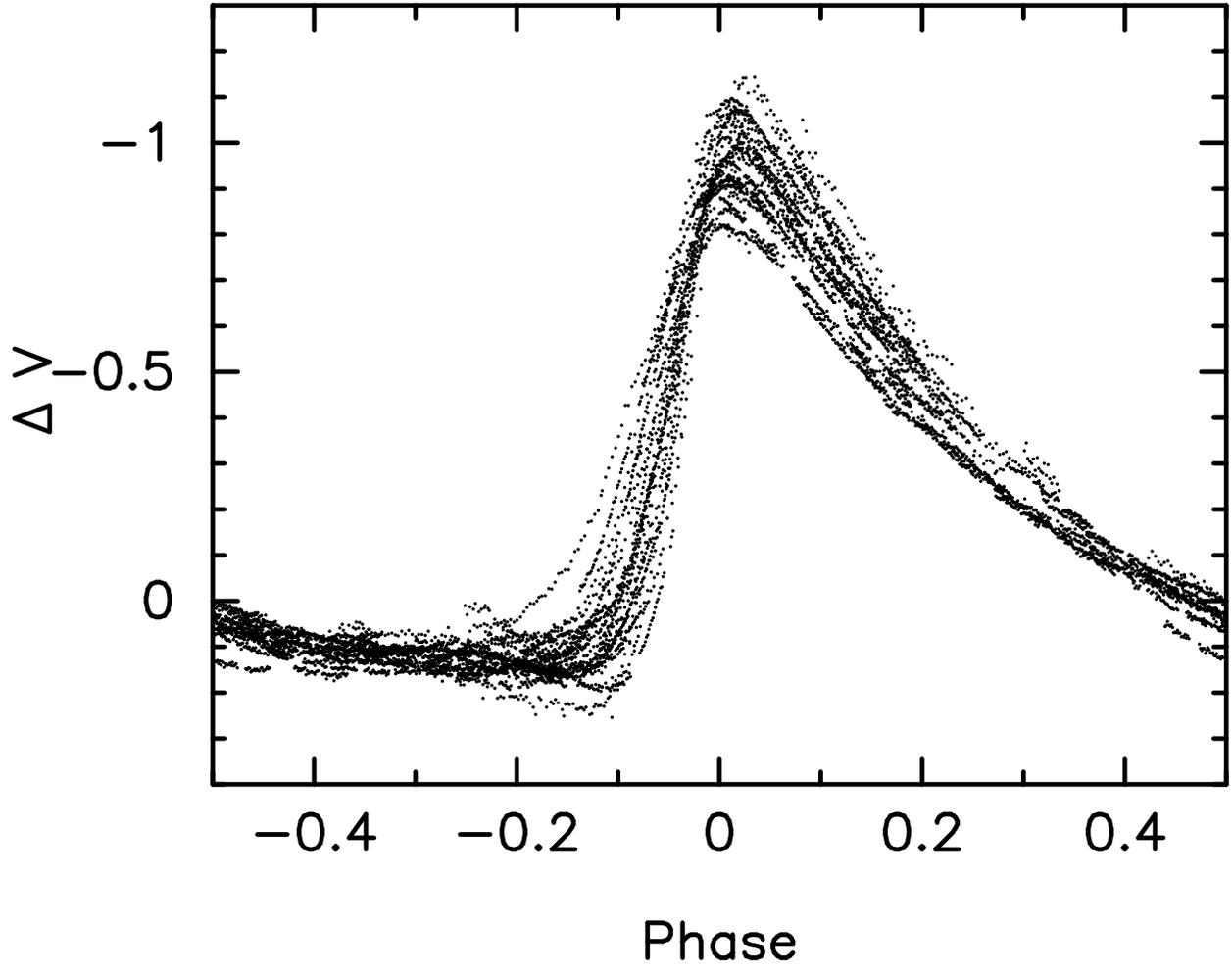}
\caption{$V$ light curve of XZ Cygni based upon
all CCD observations.} 
\end{figure} 

\clearpage 
\begin{figure}
\plotone{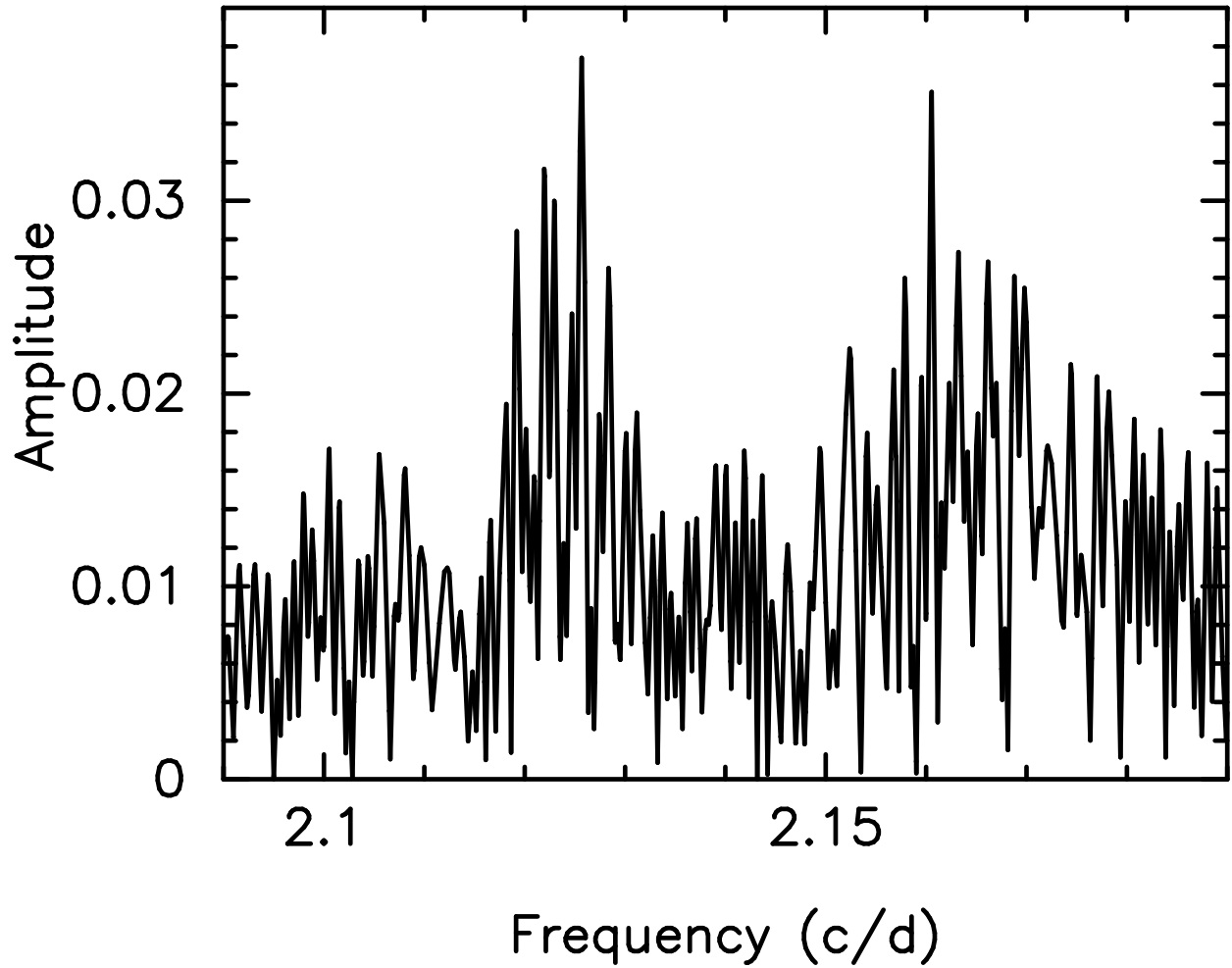}
\caption{Fourier spectrum of CCD observations of XZ Cygni,
after subtraction of the main frequency and its harmonics. Peaks
are evident at frequencies of 2.12577 c/d and 2.16054 c/d.}
\end{figure}

\clearpage 
\begin{figure}
\plotone{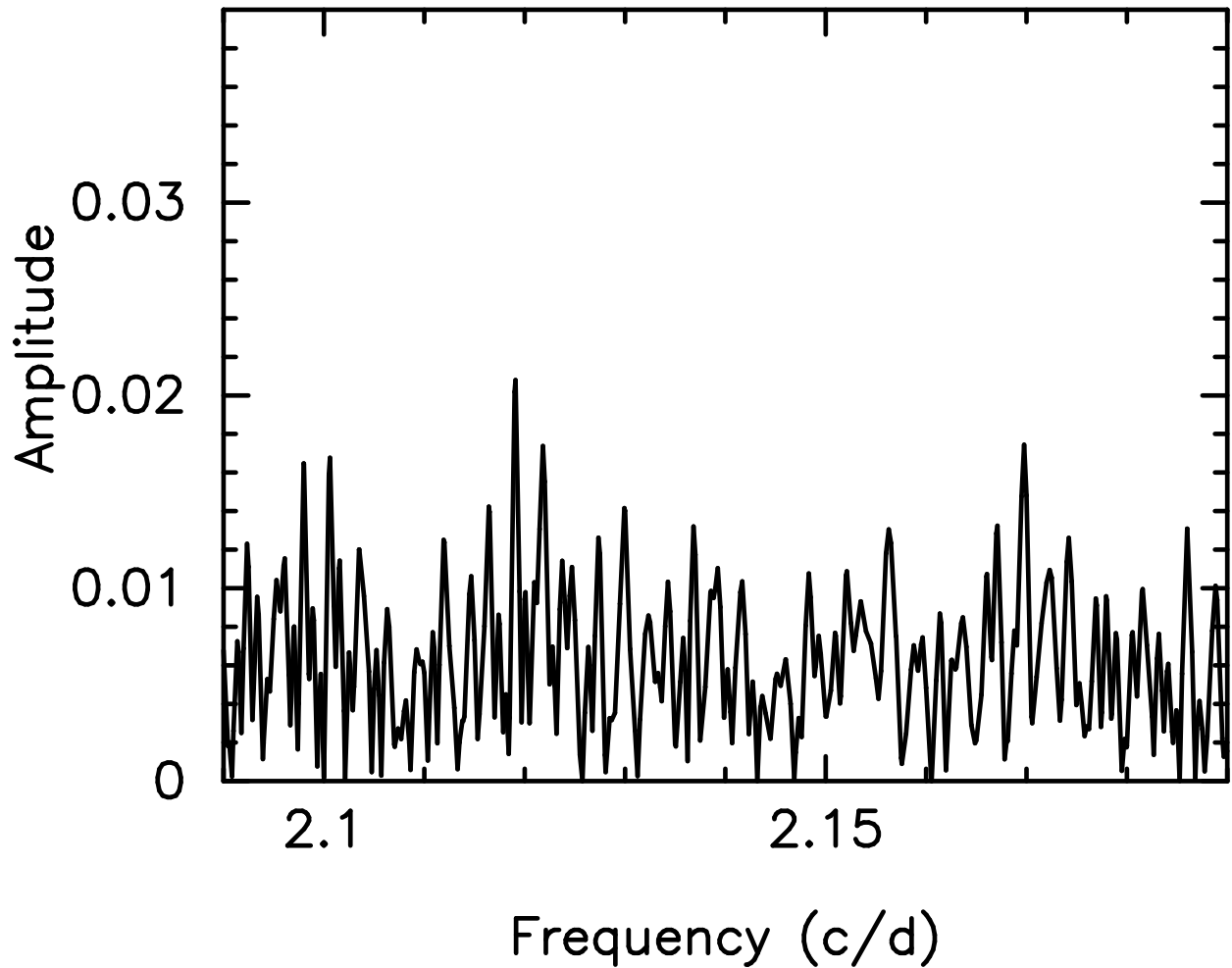}
\caption{Fourier spectrum of the CCD observations,
after subtraction of the frequency components of the primary
period and the 57.5~d Blazhko period. Peaks at 2.1191 c/d and
2.1671 c/d would beat with the primary period at intervals of
41.6 and 41.8 days. }
\end{figure}

\clearpage 
\begin{figure}
\plotone{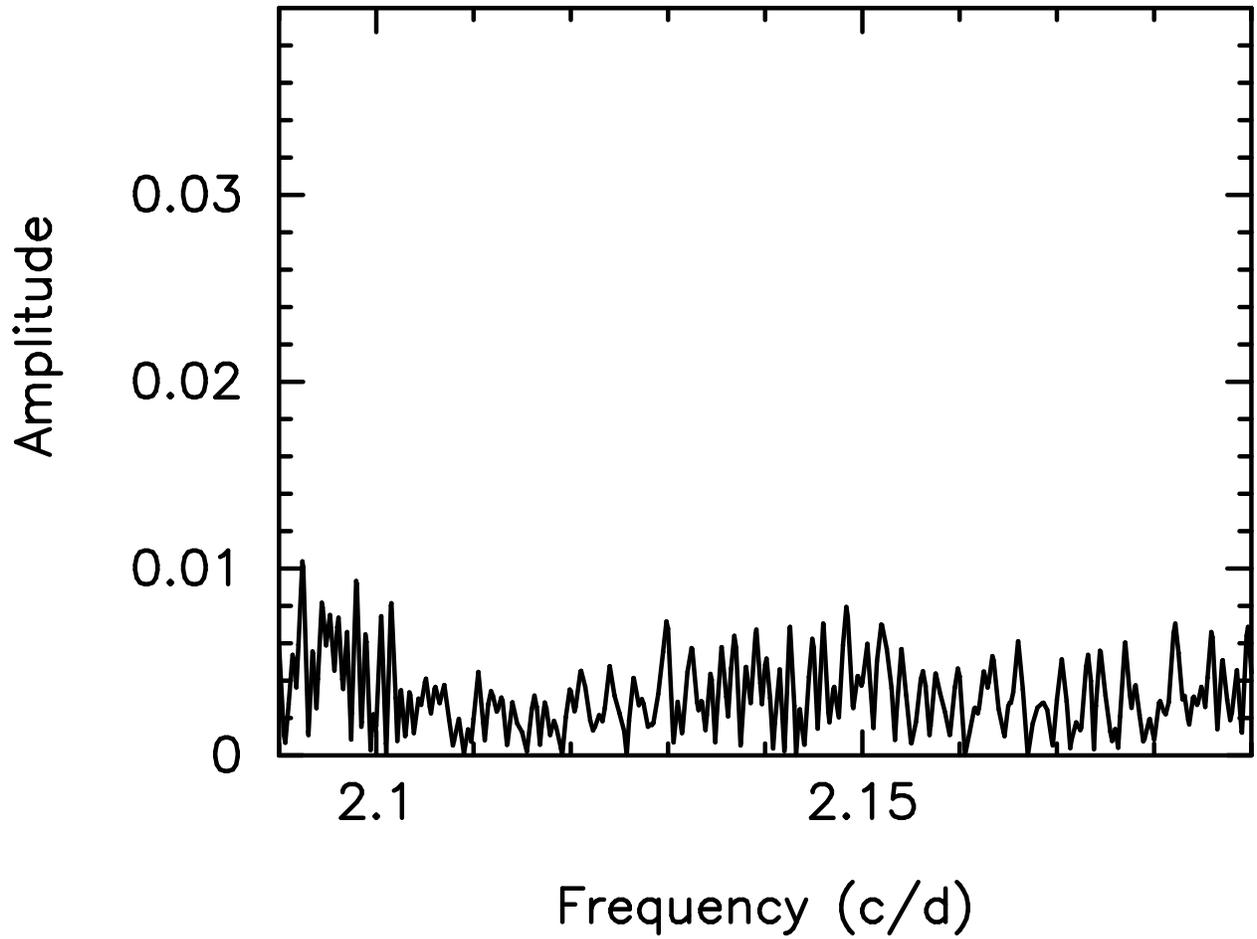}
\caption{Fourier spectrum of the CCD observations,
after subtraction of the frequency components of the primary
period, a 57.5~d period, and a 41.6~d period.}
\end{figure}

\clearpage 
\begin{figure}
\plotone{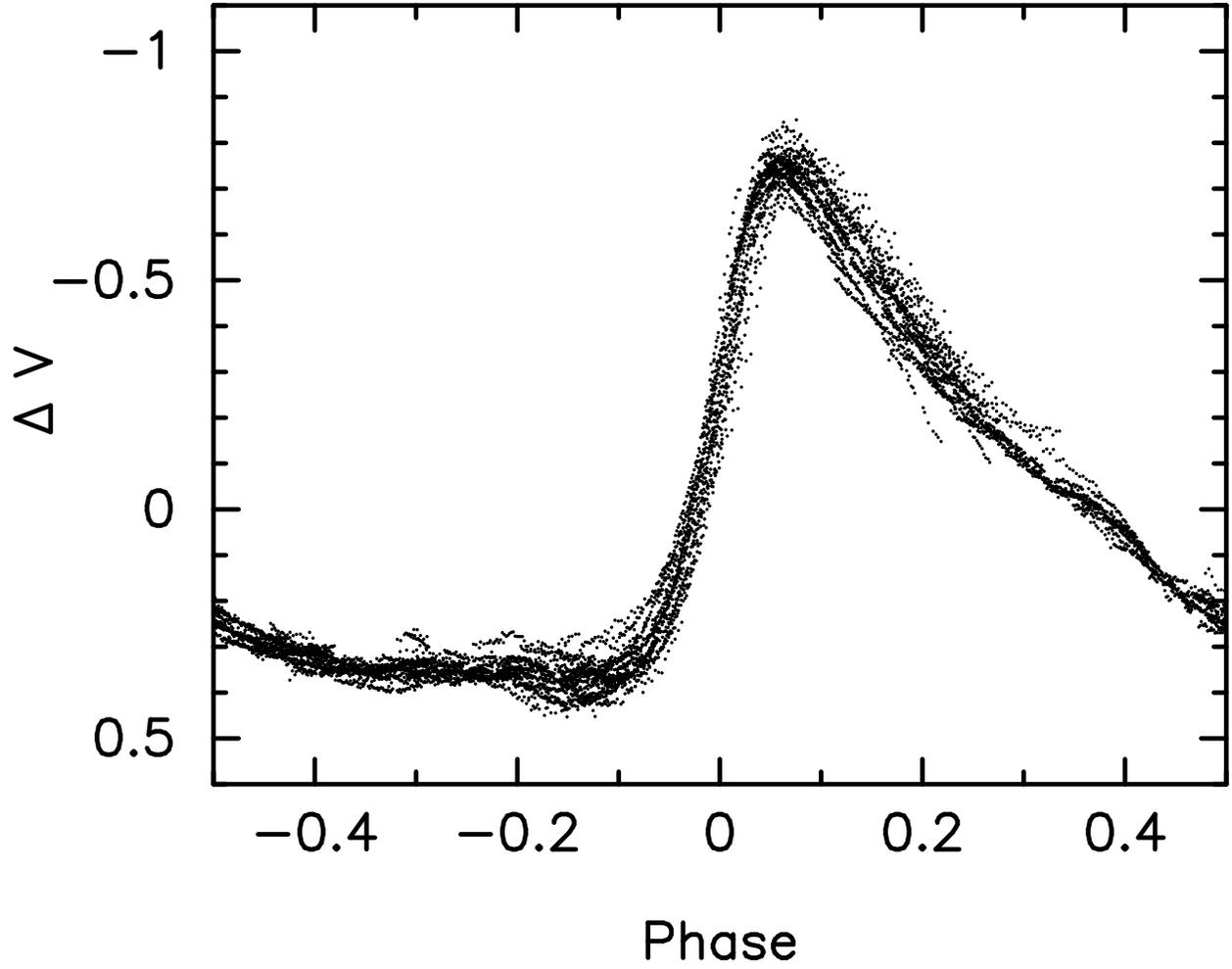}
\caption{$V$ light curve of all CCD observations, with the
sidelobes of the frequency triplets removed.}
\end{figure}

\clearpage 
\begin{figure}
\plotone{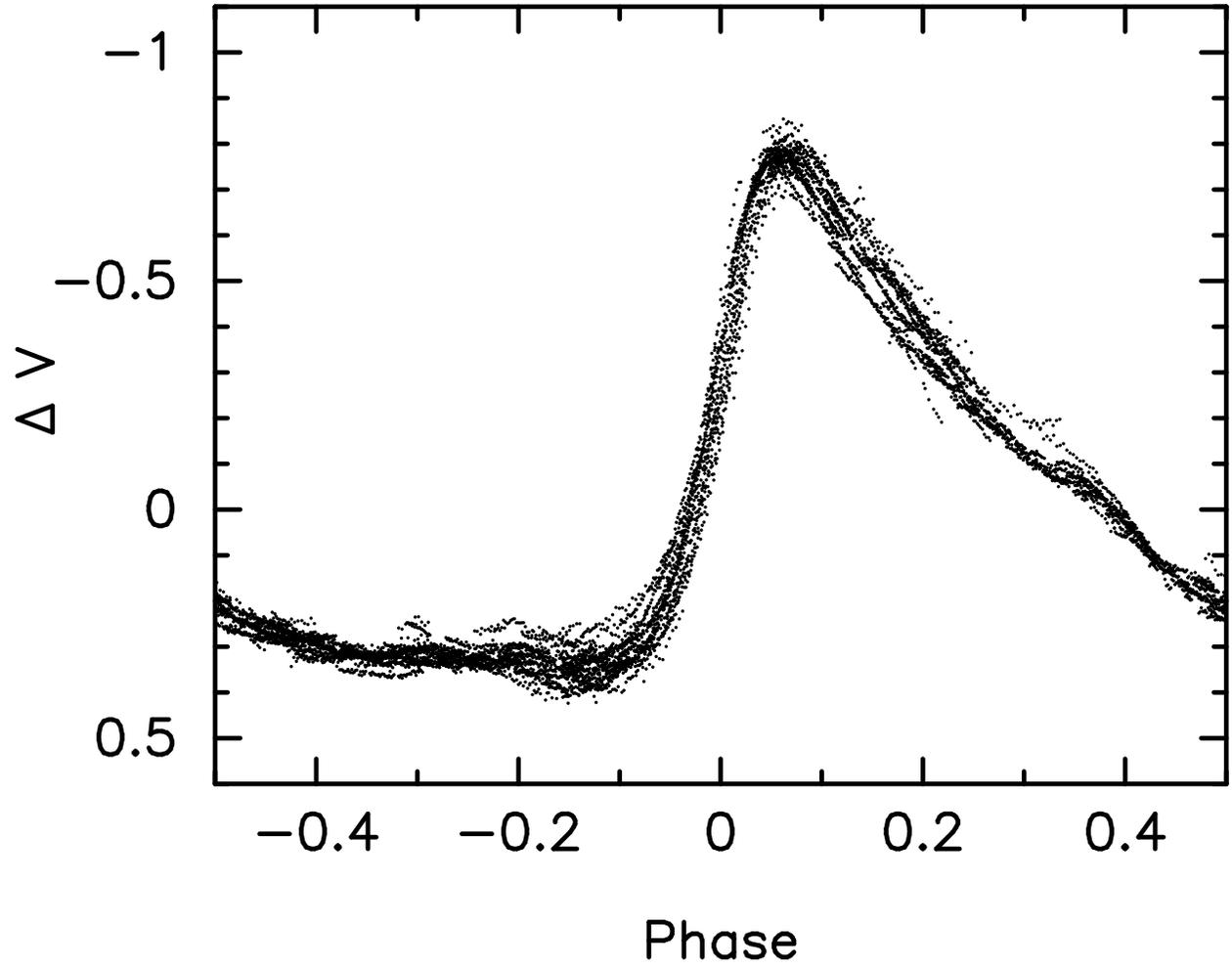}
\caption{$V$ light curve of Michigan State
University CCD observations, with the
sidelobes of the frequency triplets removed.}
\end{figure}

\clearpage 
\begin{figure}
\plotone{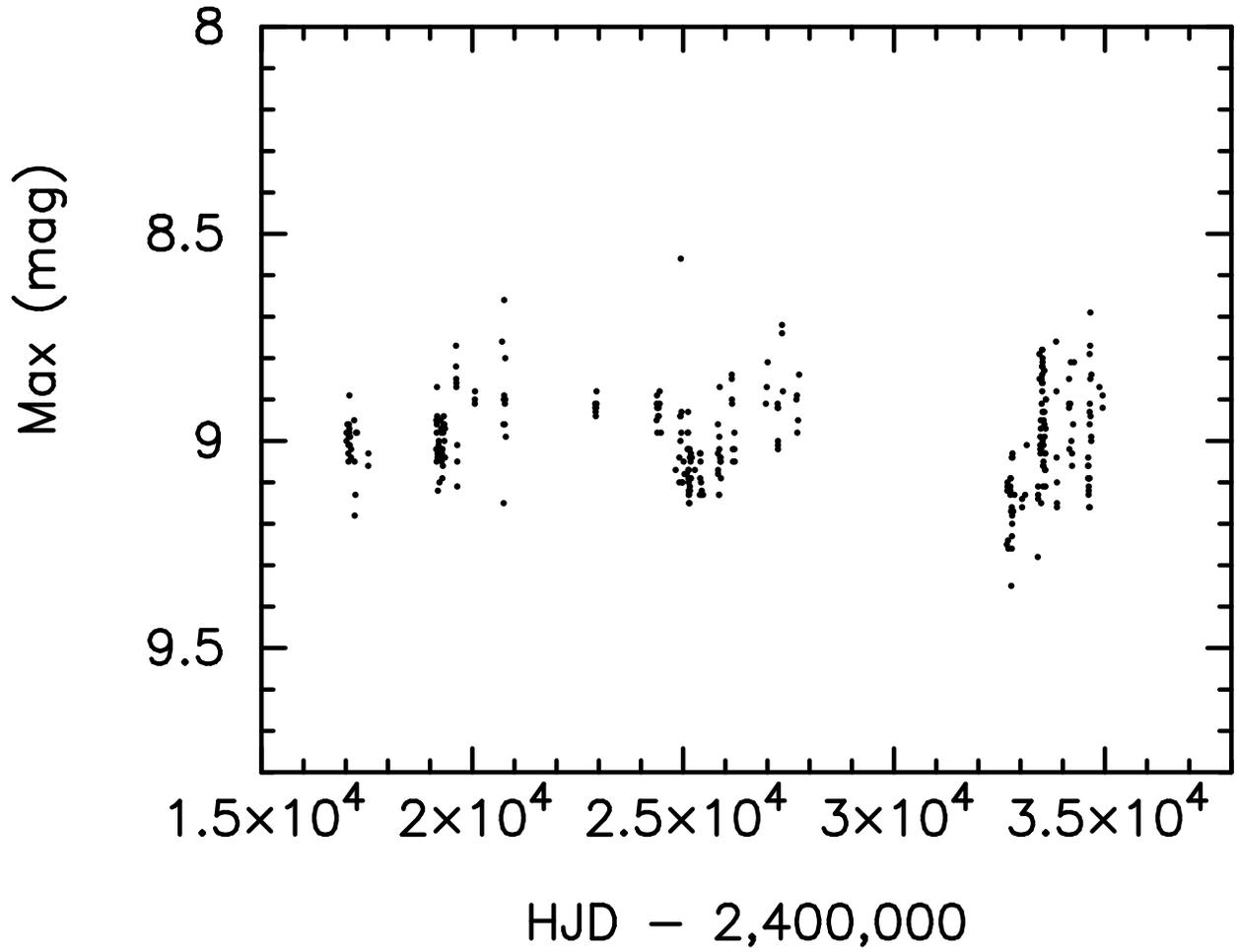}
\caption[fig8.eps]{The magnitude of XZ Cyg at maximum light versus
Julian Date for maxima listed by \citet{K58}.}
\end{figure}

\clearpage 
\begin{figure}
\plotone{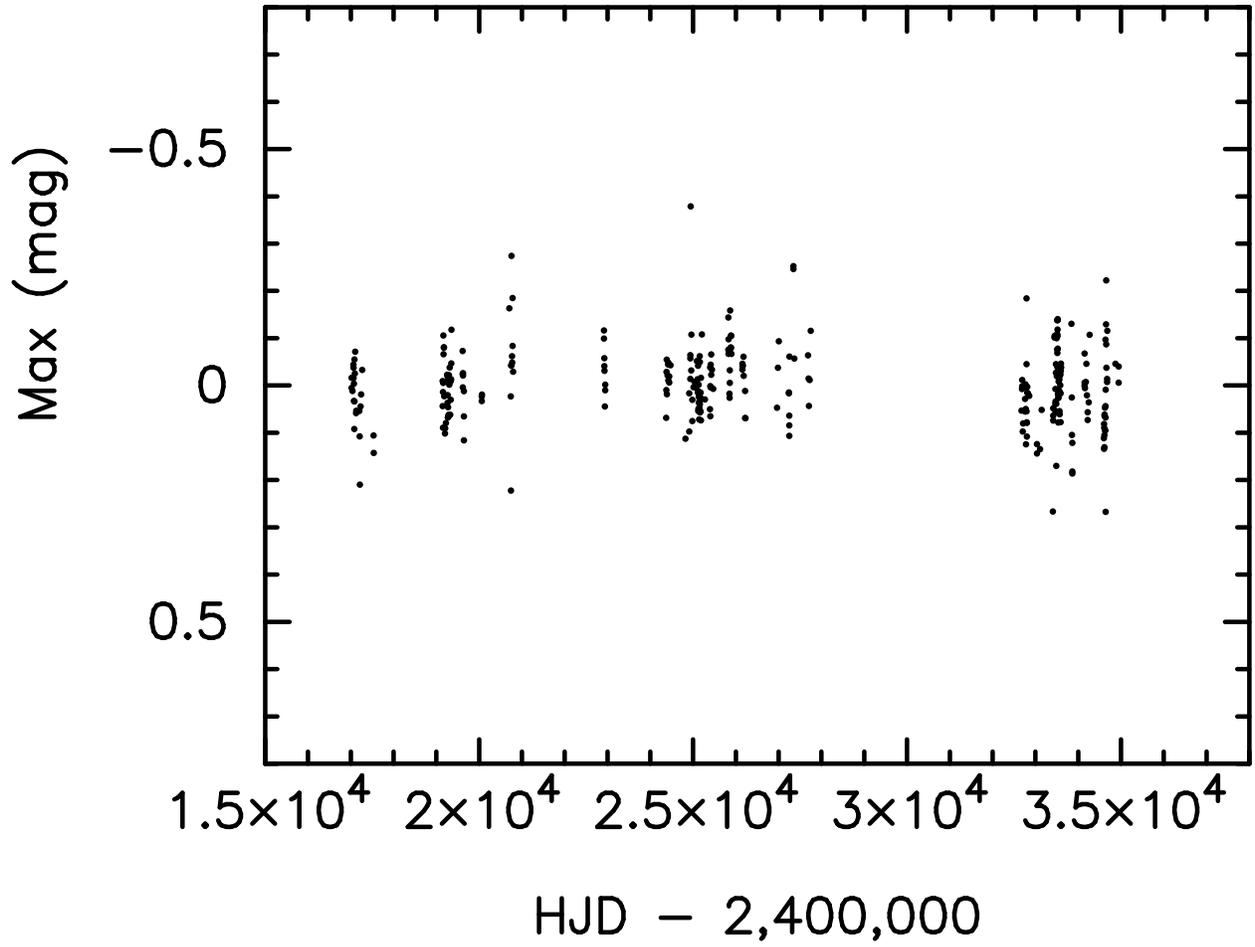}
\caption{Residuals of the magnitudes of XZ Cyg at maximum light
listed by \citet{K58} after removal of the
57.4~d, 41.6~d, and 3540~d periodicities.}
\end{figure}

\clearpage 
\begin{figure}
\plotone{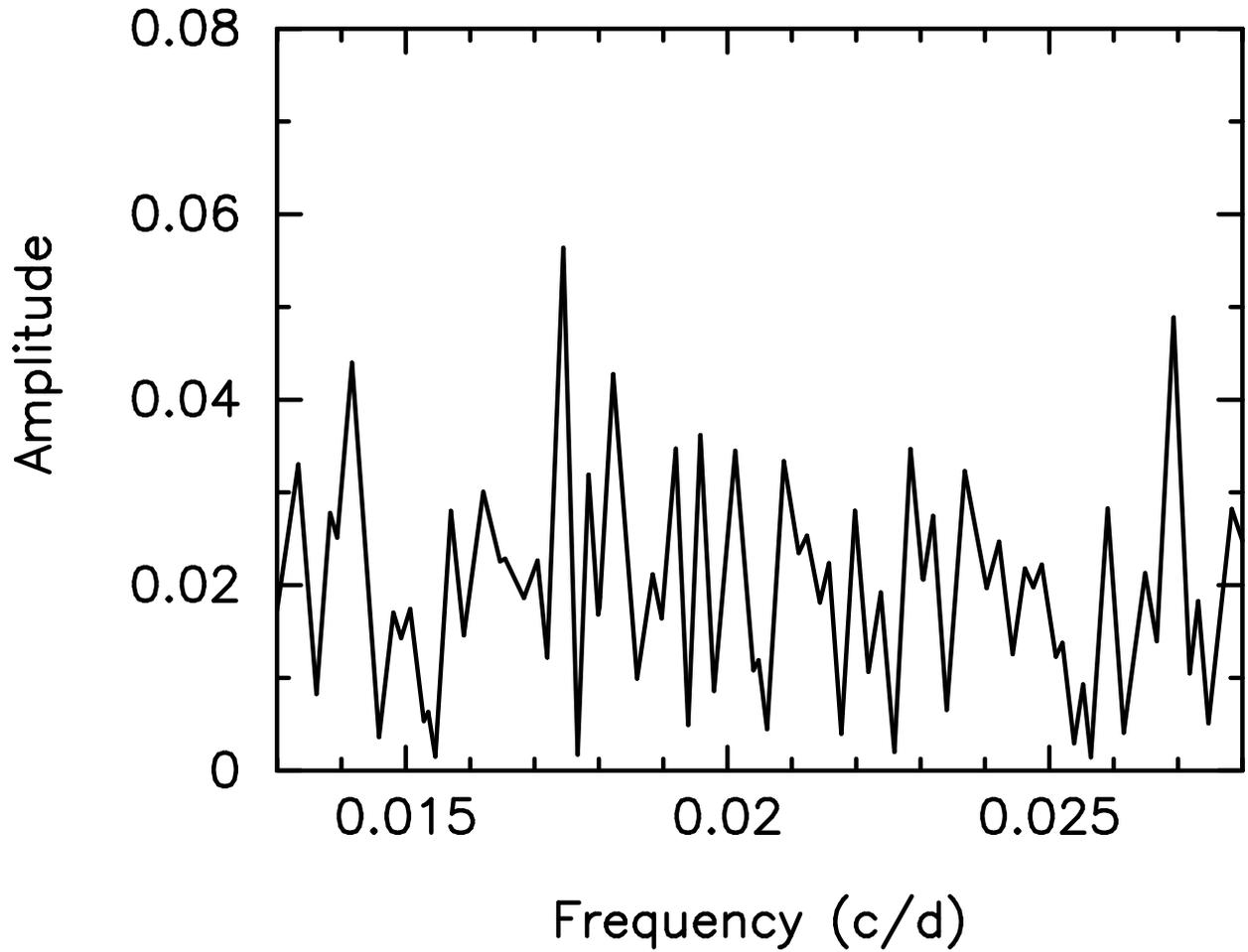}
\caption{Fourier spectrum for magnitudes at maximum for
the interval JD 2424370 to 2427750.  The peak at a frequency of
0.01744 c/d corresponds to a Blazhko period of 57.34~d.}
\end{figure}

\clearpage 
\begin{figure}
\plotone{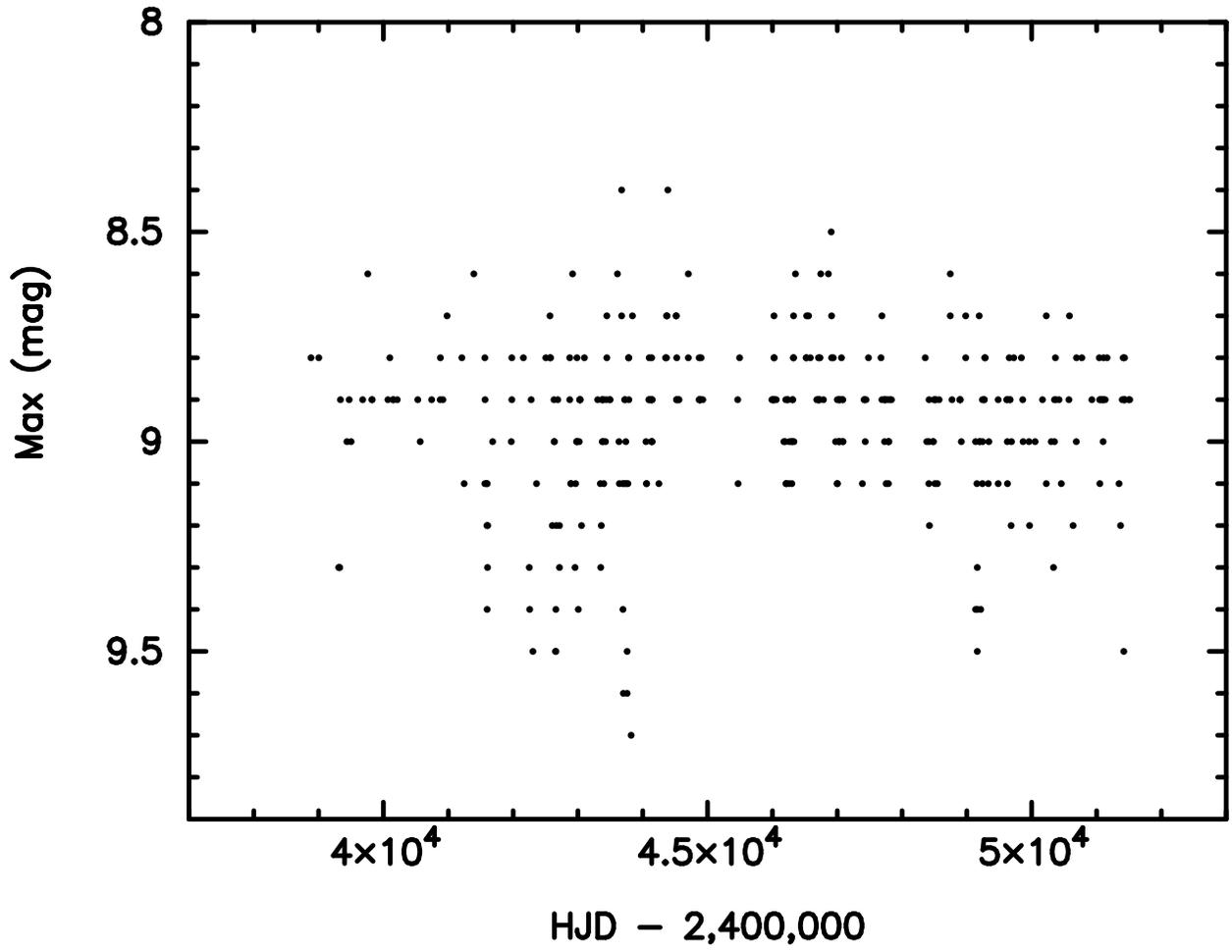}
\caption{Magnitude at maximum light versus Julian Date
for maxima observed by M. Baldwin.}
\end{figure}

\clearpage 
\begin{figure}
\plotone{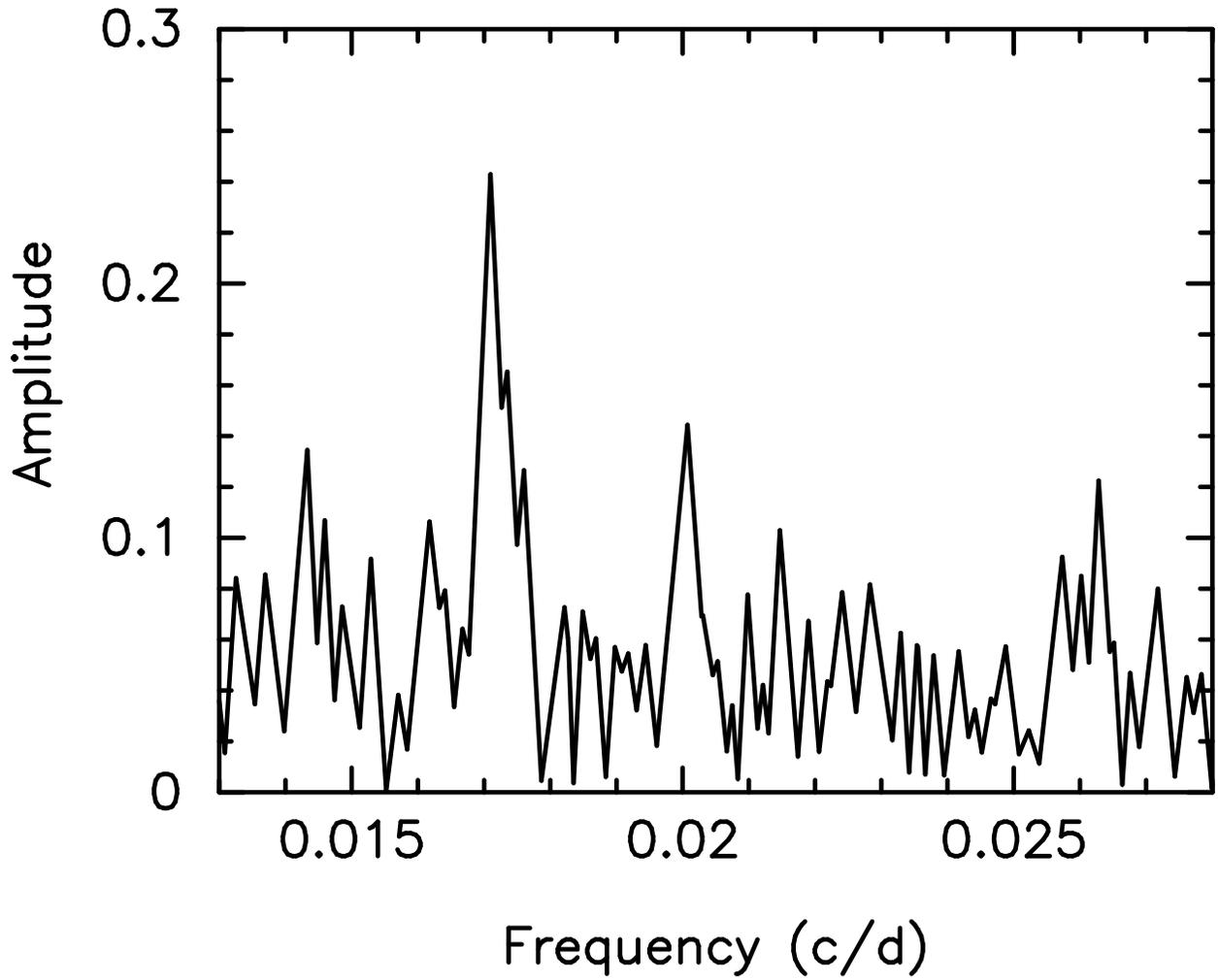}
\caption{Fourier spectrum for maximum magnitudes observed
by M. Baldwin between JD 2438882 and JD 2443800.}
\end{figure}

\clearpage 
\begin{figure}
\plotone{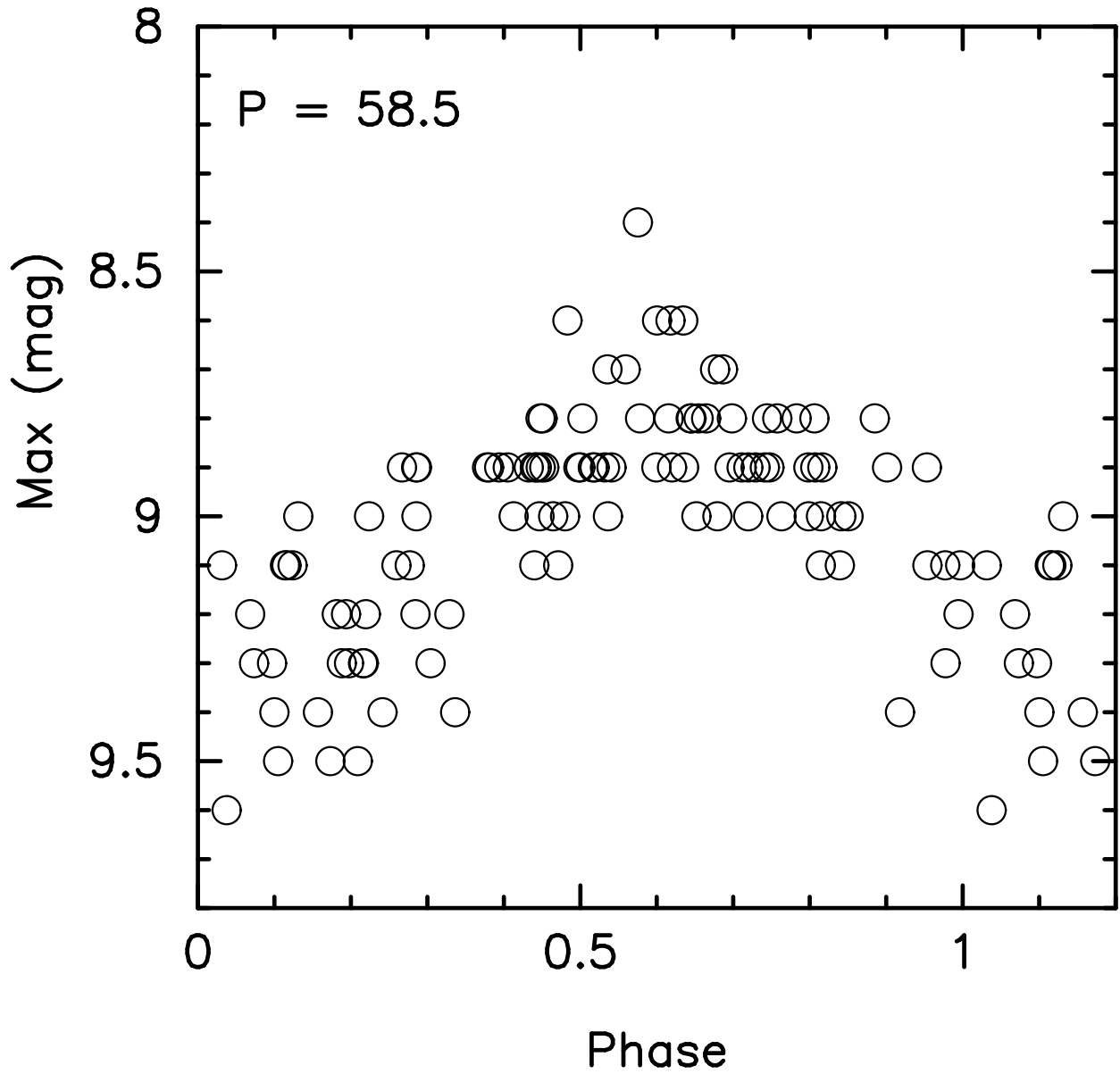}
\caption{Magnitudes of XZ Cyg at maximum light observed
by M. Baldwin between JD 2438882 and JD 2443800, folded with a 58.5~d
period.}
\end{figure}

\clearpage 
\begin{figure}
\plotone{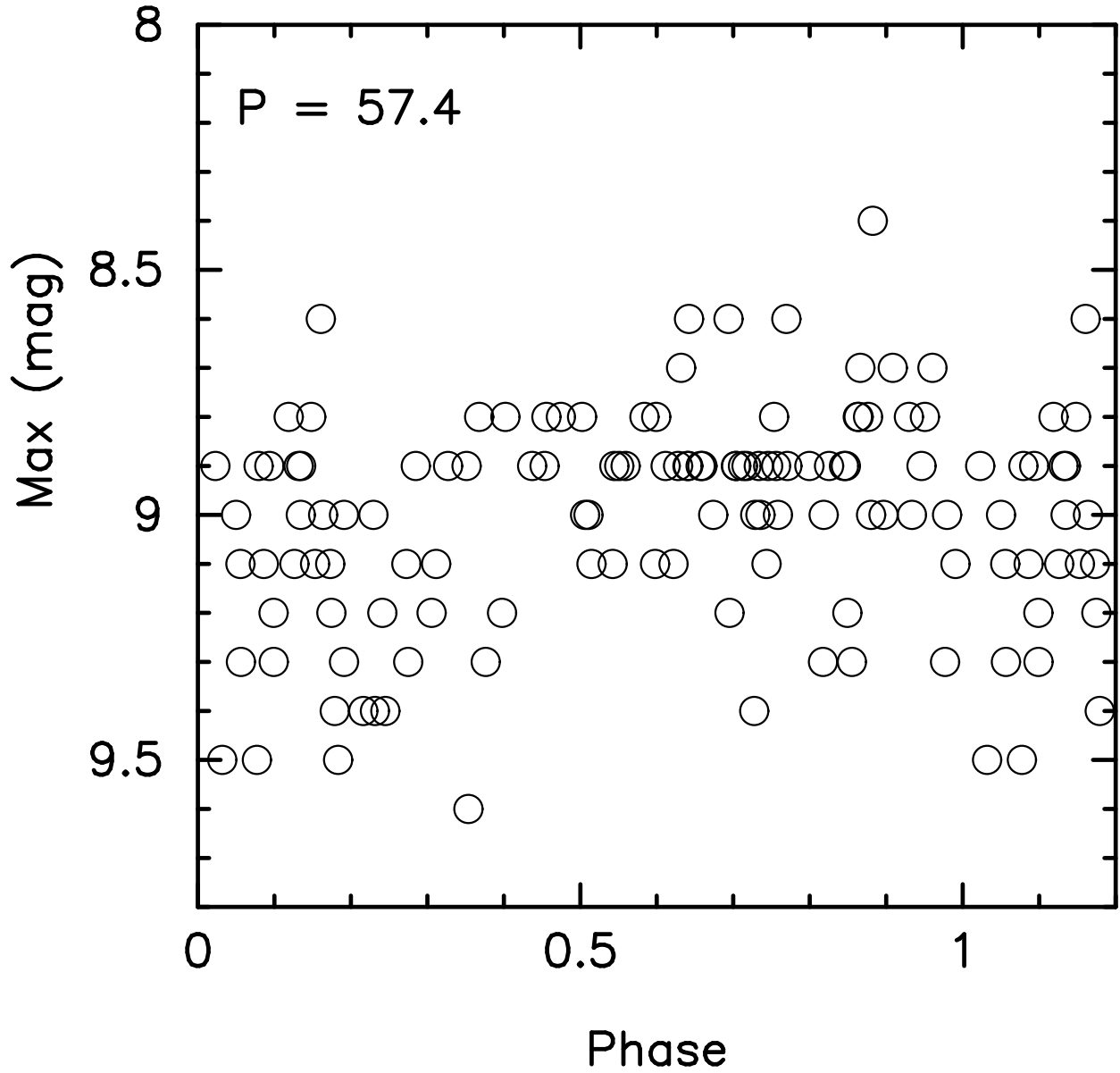}
\caption{Magnitudes of XZ Cyg at maximum light observed
by M. Baldwin between JD 2438882 and JD 2443800, folded with a 57.4~d
period. More scatter exists than in Figure 13.}
\end{figure}

\clearpage 
\begin{figure}
\plotone{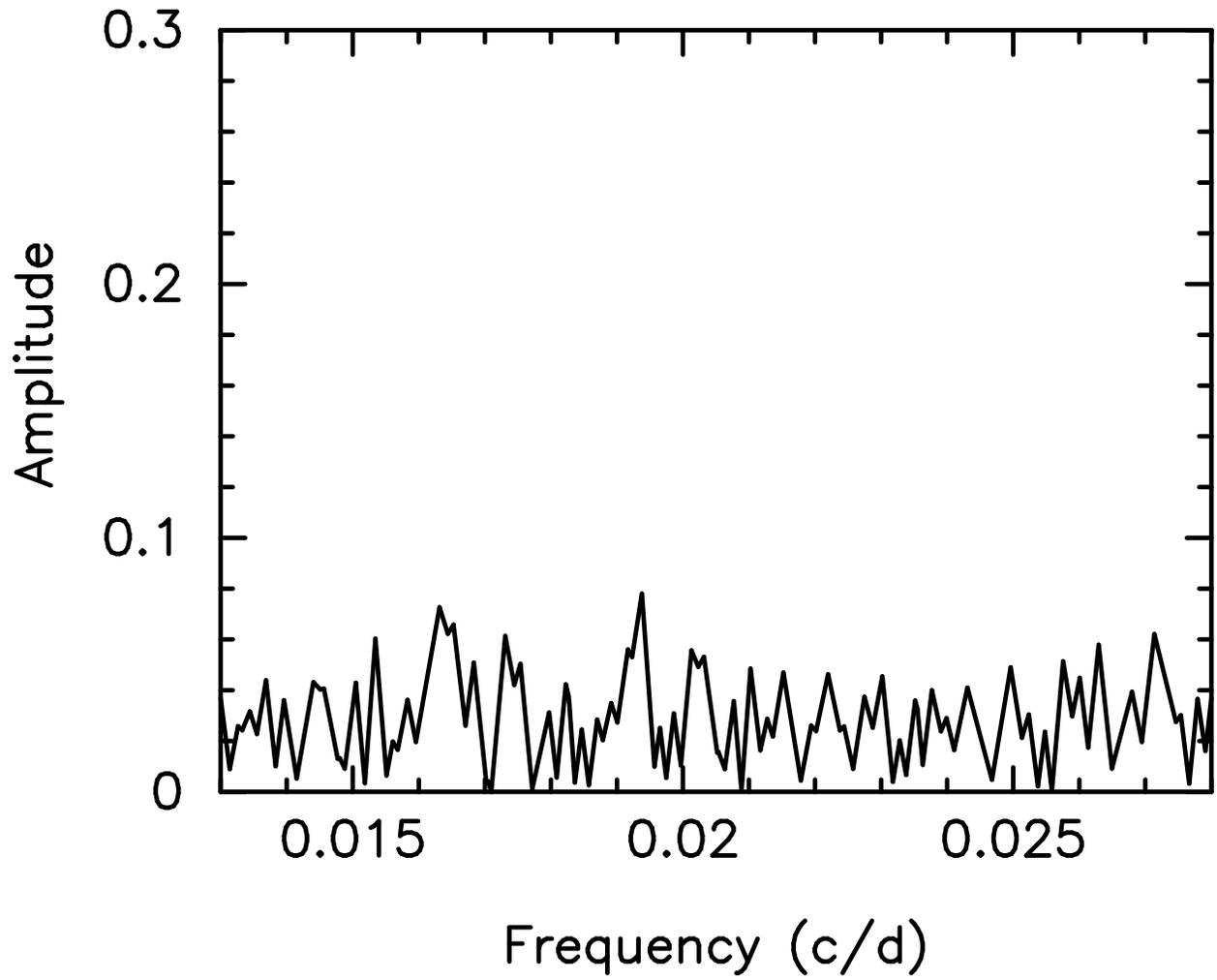}
\caption{Fourier spectrum for maximum magnitudes observed
by M. Baldwin between JD 2438882 and JD 2443800. The observations
have been prewhitened to remove the 58.5~d periodicity. }
\end{figure}

\clearpage 
\begin{figure}
\plotone{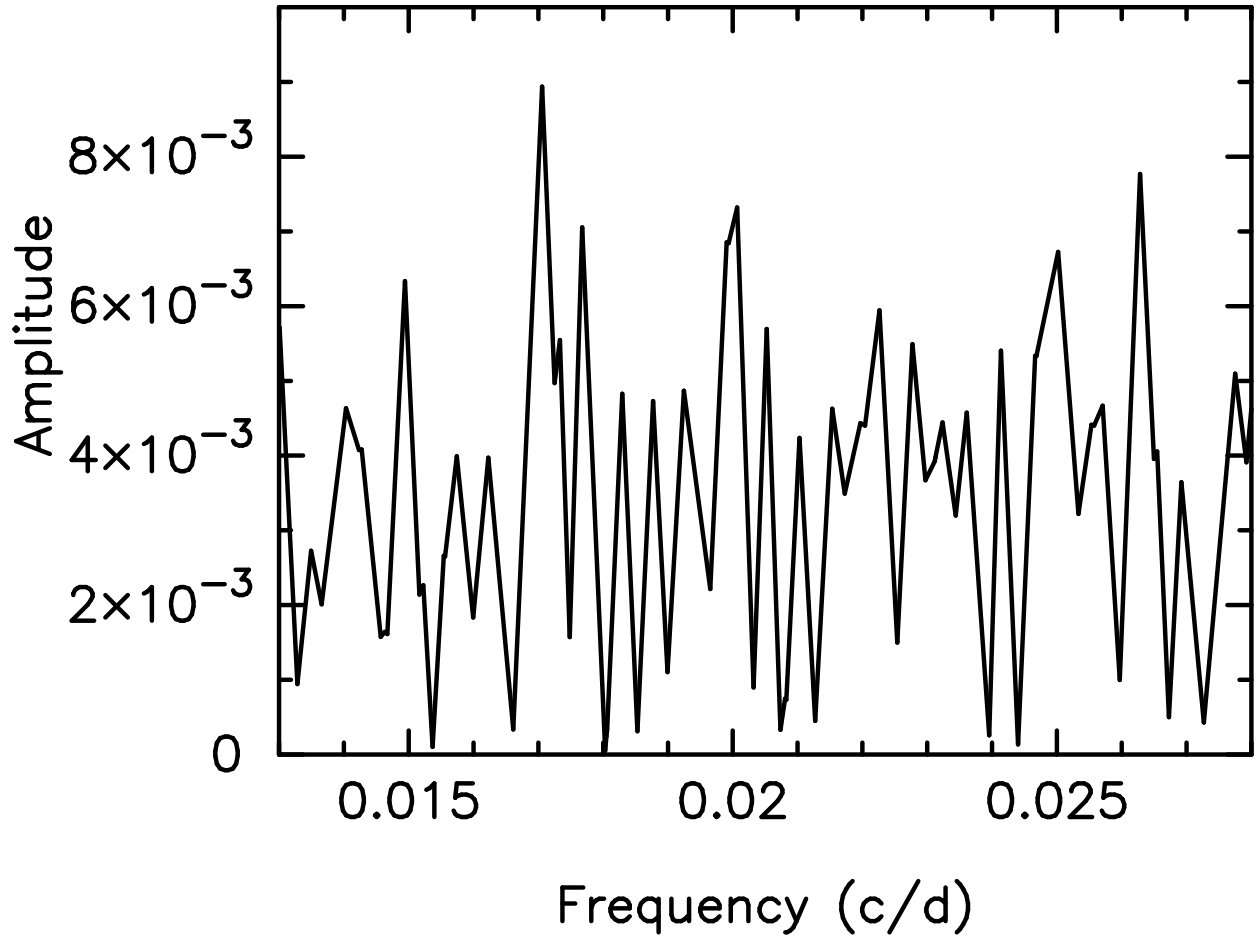}
\caption{Fourier spectrum of O-C values for the time
of maximum light, from observations obtained between 
JD 2438882 and JD 2443800. The peak at 0.01706 c/d corresponds
to a period of 58.6~d.}
\end{figure}

\clearpage 
\begin{figure}
\plotone{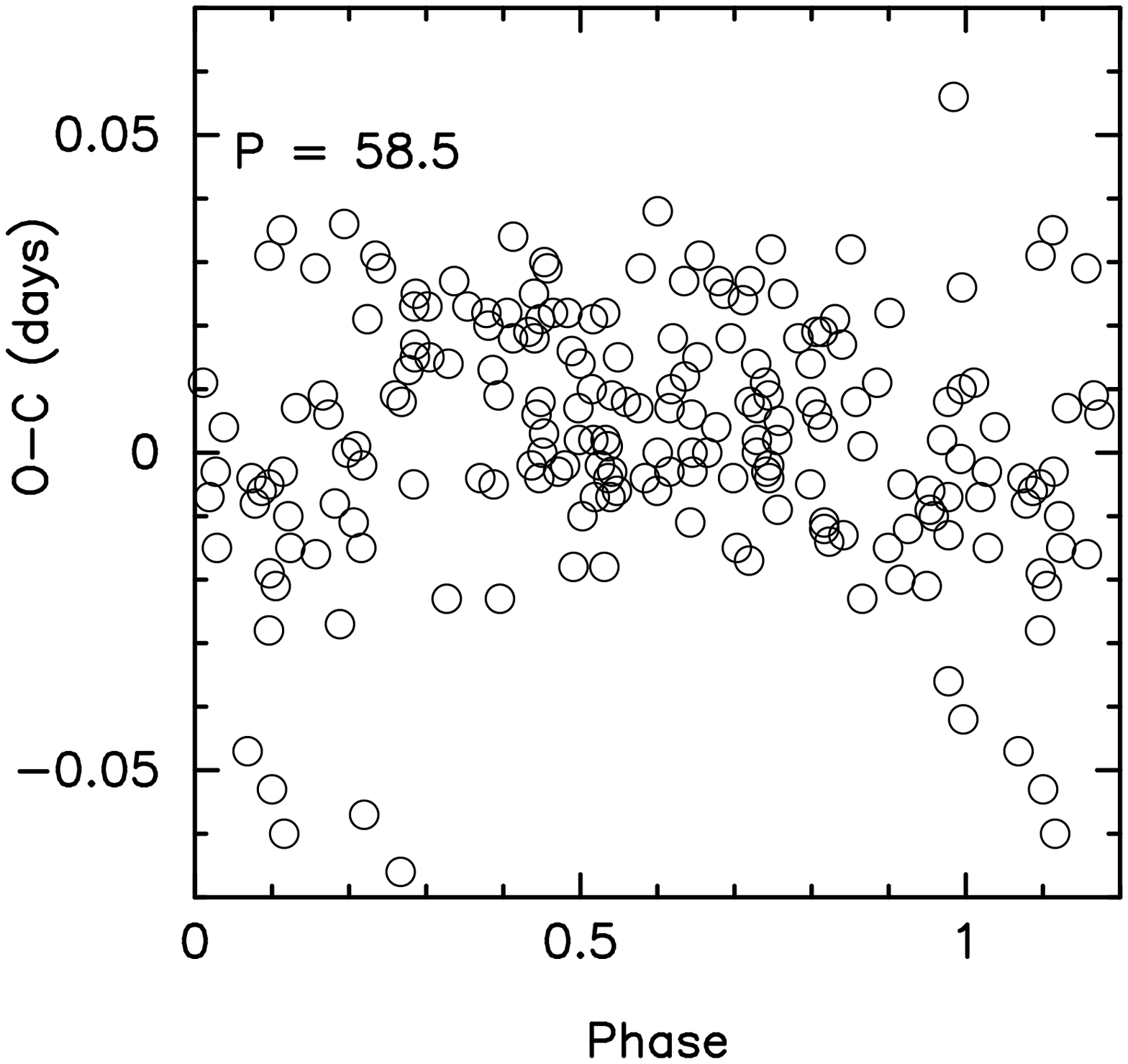}
\caption{O-C values obtained from observations made
between JD 2438882 and JD 2443800 have been folded with a period of
58.5~d. }
\end{figure}

\clearpage 
\begin{figure}
\plotone{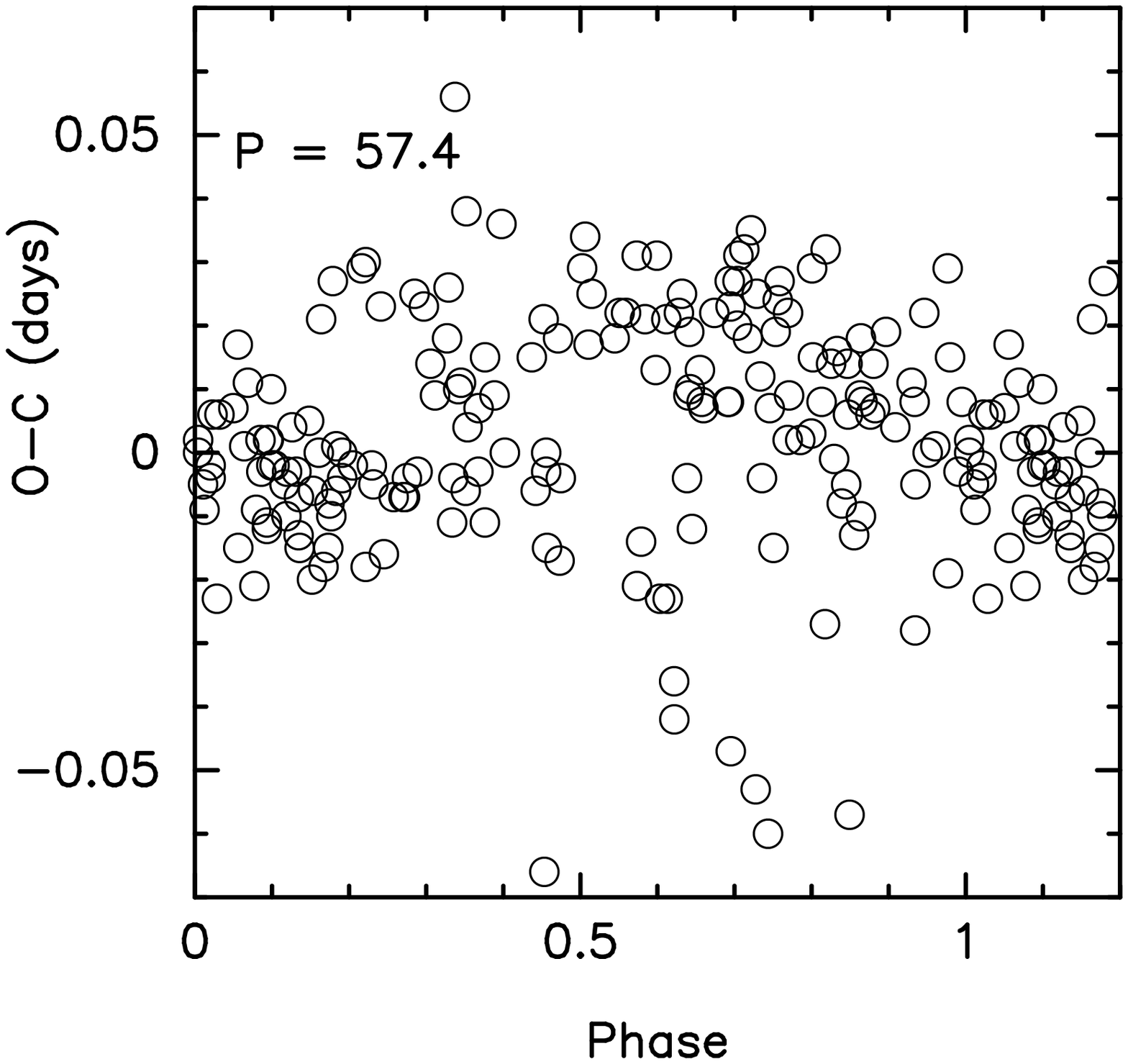}
\caption{O-C values obtained from observations made
between JD 2438882 and JD 2443800 have been folded with a period
of 57.4~d. }
\end{figure}

\clearpage 
\begin{figure}
\plotone{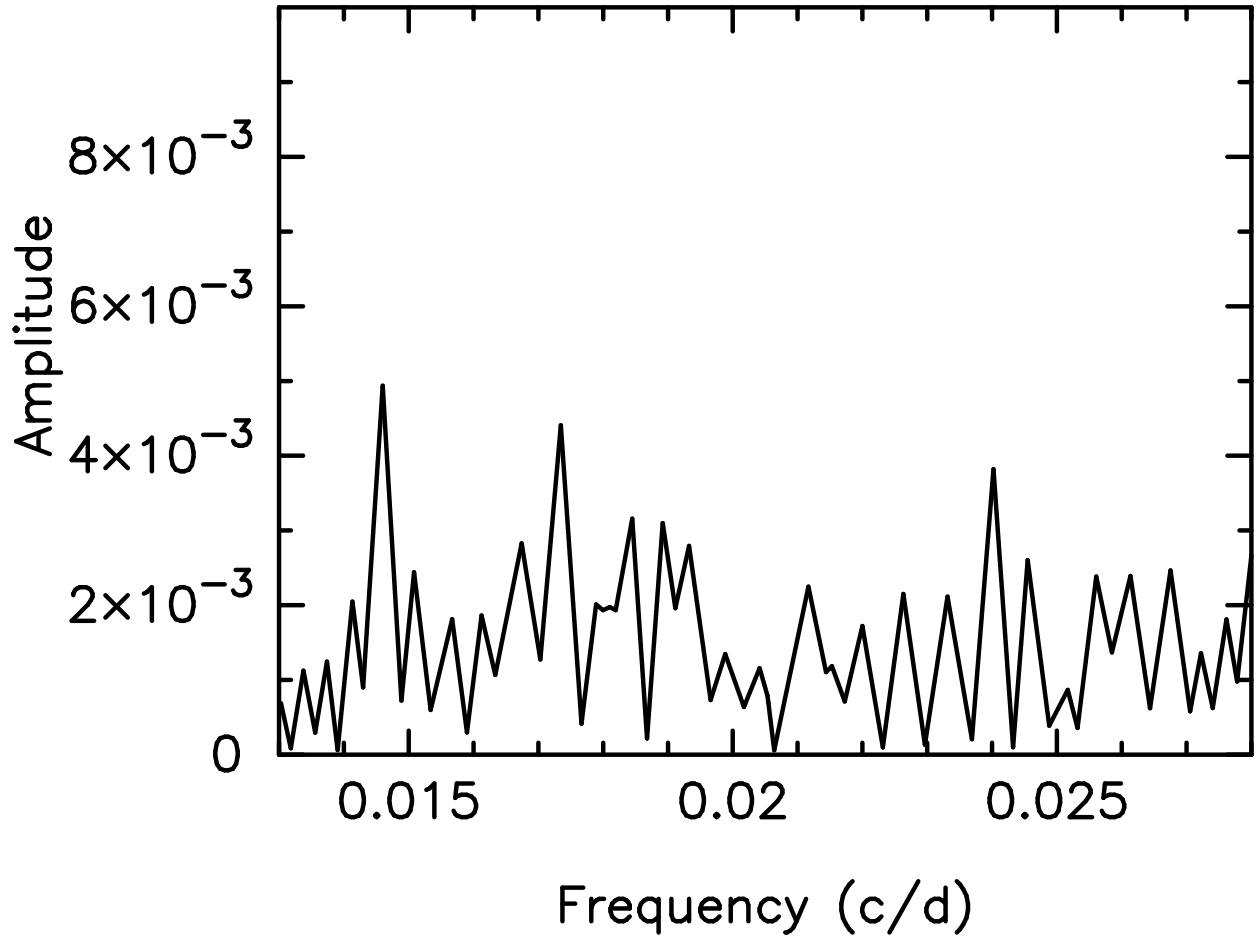}
\caption{Fourier spectrum of O-C values for the time
of maximum light, from observations obtained between 
JD 2448570 and 2452618.  Peaks at frequencies 0.01736 c/d and
0.02404 c/d correspond to periods of 57.6~d and 41.6~d.}
\end{figure}

\clearpage 
\begin{figure}
\plotone{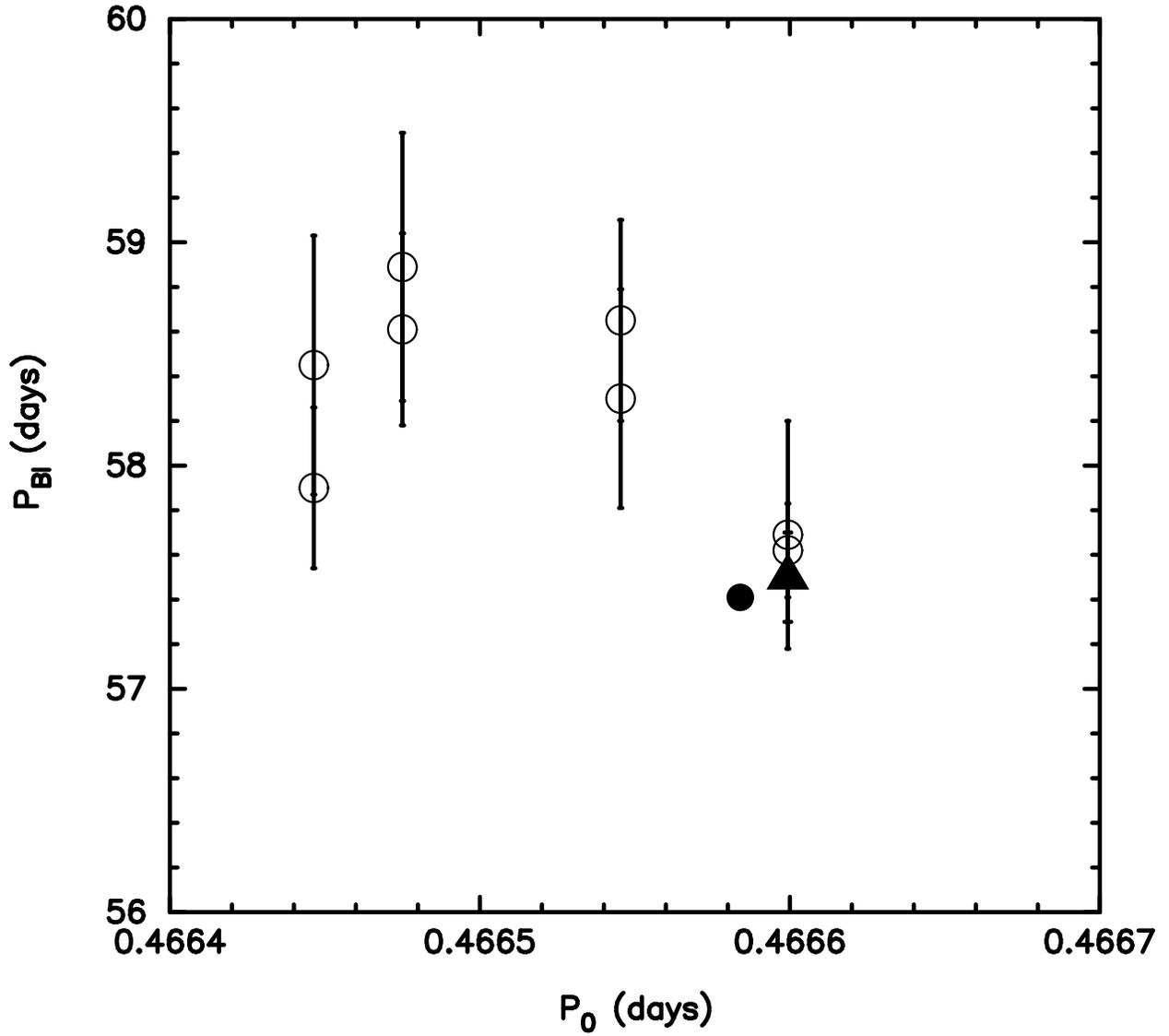}
\caption{Blazhko period plotted against the period
of the fundamental mode.  The filled triangle is based upon the
CCD observations reported in this paper.  The open circles are
taken from Tables 5 and 6.  The result for the analysis of the
maxima listed by \citet{K58} is also included (filled
circle). }
\end{figure}

\clearpage
\begin{deluxetable}{cc} 
\tablewidth{0pc}
\footnotesize
\tablenum{1}
\tablecaption{CCD Photometry (Michigan State University)}
\tablehead{
\colhead{HJD $-$ 2450000} & \colhead{$\Delta$V}
          }
\startdata
1338.6430 & $-$0.004 \\  
1338.6452 & $-$0.027\\
1338.6470 & $-$0.028 \\
1338.6487 & $-$0.019 \\
\enddata
\tablecomments{The complete version of this table is in the electronic 
edition of the Journal.  The printed edition contains only a sample.}
\end{deluxetable}

\clearpage

\begin{deluxetable}{cc} 
\tablewidth{0pc}
\footnotesize
\tablenum{2}
\tablecaption{CCD Photometry (Milwaukee)}
\tablehead{
\colhead{HJD $-$ 2450000} & \colhead{$\Delta$V}
          }
\startdata
2210.49663  & $-$0.675 \\
2210.49810  & $-$0.693 \\
2210.49957  & $-$0.721 \\
2210.50103  & $-$0.735 \\
\enddata
\tablecomments{The complete version of this table is in the electronic 
edition of the Journal.  The printed edition contains only a sample.}
\end{deluxetable}
\clearpage

\begin{deluxetable} {cccl}
\tablenum{3}
\footnotesize
\tablecaption{Parameters for fit to CCD data \label{tbl-3}}
\tablewidth{0pt}  
\tablehead{
\colhead{Frequency $(d^{-1})$} & \colhead{Amplitude (mag)} & \colhead{Phase (cycles)}
& \colhead{ID}}
\startdata
2.14317&0.447&0.090&$f_0$ \\
2.16056&0.031&0.708&$f_0$ + $f_B$ \\
2.12578&0.027&0.267&$f_0$ $-$ $f_B$ \\
2.16721&0.026&0.824&$f_0$ + $f_3$ \\
2.11913&0.027&0.613&$f_0$ $-$ $f_3$ \\
\\
4.28634&0.186&0.632&$2f_0$ \\
4.30374&0.022&0.242&$2f_0$ + $f_B$ \\
4.26895&0.012&0.656&$2f_0$ $-$ $f_B$ \\
4.31038&0.013&0.795&$2f_0$ + $f_3$ \\
4.26231&0.015&0.233&$2f_0$ $-$ $f_3$ \\
\\
6.42952&0.126&0.172&$3f_0$ \\
6.44691&0.017&0.906&$3f_0$ + $f_B$ \\
6.41213&0.019&0.394&$3f_0$ $-$ $f_B$ \\
6.45356&0.009&0.697&$3f_0$ + $f_3$ \\
6.40548&0.020&0.679&$3f_0$ $-$ $f_3$ \\
\\
8.57269&0.070&0.626&$4f_0$ \\
8.59008&0.020&0.161&$4f_0$ + $f_B$ \\
8.55530&0.015&0.483&$4f_0$ $-$ $f_B$ \\
8.59673&0.014&0.119&$4f_0$ + $f_3$ \\
8.54865&0.016&0.867&$4f_0$ $-$ $f_3$ \\
\\
10.71586&0.037&0.292&$5f_0$ \\
10.73325&0.009&0.845&$5f_0$ + $f_B$ \\
10.69847&0.008&0.099&$5f_0$ $-$ $f_B$ \\
10.73990&0.009&0.852&$5f_0$ + $f_3$ \\
10.69182&0.012&0.742&$5f_0$ $-$ $f_3$ \\
\\
12.85904&0.027&0.533&$6f_0$ \\
12.87643&0.010&0.136&$6f_0$ + $f_B$ \\
12.84164&0.008&0.359&$6f_0$ $-$ $f_B$ \\
12.88307&0.009&0.297&$6f_0$ + $f_3$ \\
12.83500&0.008&0.072&$6f_0$ $-$ $f_3$ \\
\\
15.00221&0.015&0.170&$7f_0$ \\
15.01960&0.007&0.758&$7f_0$ + $f_B$ \\
14.98482&0.005&0.653&$7f_0$ $-$ $f_B$ \\
15.02625&0.003&0.312&$7f_0$ + $f_3$ \\
14.97817&0.004&0.542&$7f_0$ $-$ $f_3$ \\
\\
17.14538&0.011&0.981&$8f_0$ \\
17.16277&0.005&0.082&$8f_0$ + $f_B$ \\
17.12799&0.006&0.261&$8f_0$ $-$ $f_B$ \\
17.16942&0.001&0.397&$8f_0$ + $f_3$ \\
17.12134&0.003&0.914&$8f_0$ $-$ $f_3$ \\
\\
19.28855&0.010&0.927&$9f_0$ \\
19.30594&0.007&0.567&$9f_0$ + $f_B$ \\
19.28855&0.004&0.853&$9f_0$ $-$ $f_B$ \\
19.31259&0.002&0.468&$9f_0$ + $f_3$ \\
19.26451&0.003&0.569&$9f_0$ $-$ $f_3$ \\
\\
21.43137&0.004&0.307&$10f_0$ \\
21.44876&0.005&0.795&$10f_0$ + $f_B$ \\
21.41398&0.002&0.806&$10f_0$ $-$ $f_B$ \\
21.45576&0.002&0.619&$10f_0$ + $f_3$ \\
21.40767&0.004&0.933&$10f_0$ $-$ $f_3$ \\

\enddata
\end{deluxetable}
\clearpage

\begin{deluxetable}{cl} 
\tablewidth{0pc}
\footnotesize
\tablenum{4}
\tablecaption{XZ Cyg primary period (Baldwin \& Samolyk 2003)}
\tablehead{
\colhead{Julian Date} & \colhead{Primary period (days)}
          }
\startdata
2438880 $-$ 2440520  & 0.4665454 \\
2440520 $-$ 2442050  & 0.4664750 \\
2442050 $-$ 2443740  & 0.4664464 \\
2444050 $-$ 2445270  & 0.4666938 \\
2445270 $-$ 2446440 & 0.4666263 \\
2446440 $-$ 2448570 & 0.46661795 \\
2448570 $-$ 2452618 & 0.46659934 \\
\enddata
\end{deluxetable}
\clearpage

\begin{deluxetable}{cccc} 
\tablewidth{0pc}
\footnotesize
\tablenum{5}
\tablecaption{Blazhko Period based on Maximum Mag}
\tablehead{
\colhead{Julian Date} & \colhead{Blazhko Period (days)} & \colhead{Amp.(mag)}
& \colhead{Nobs}
          }
\startdata
2438882 $-$ 2440529   & 58.65 $\pm$ 0.45  & 0.15 $\pm$ 0.06 & 31  \\
2440559 $-$ 2442050   & 58.89 $\pm$ 0.60 & 0.26 $\pm$ 0.08 & 22  \\
2442050 $-$ 2443740  & 57.90  $\pm$ 0.36  & 0.28 $\pm$ 0.06 & 68 \\
2444049 $-$ 2445270  & No clear period  & - & 37 \\
2445270 $-$ 2446440 & No clear period & - & 31  \\
2446440 $-$ 2448570 & No clear period & - & 68 \\
2448570 $-$ 2451510 & 57.69 $\pm$ 0.51 & 0.07 $\pm$ 0.05 & 87 \\
\enddata
\end{deluxetable}
\clearpage

\begin{deluxetable}{cccc} 
\tablewidth{0pc}
\footnotesize
\tablenum{6}
\tablecaption{Blazhko Period based on O-C}
\tablehead{
\colhead{Julian Date} & \colhead{Blazhko Period (days)} & \colhead{Amp(days)}
& \colhead{Nobs}
          }
\startdata
2438882 $-$ 2440529   & 58.30 $\pm$ 0.49  & 0.011 $\pm$ 0.005 & 31  \\
2440559 $-$ 2442050   & 58.61 $\pm$ 0.43 & 0.018 $\pm$ 0.011 & 53  \\
2442050 $-$ 2443848  & 58.45  $\pm$ 0.58  & 0.008 $\pm$ 0.004 & 96 \\
2444049 $-$ 2445270  & No clear period & - & 119 \\
2445270 $-$ 2446440 & No clear period & - & 55 \\
2446440 $-$ 2448570 & No clear period & - & 91 \\
2448570 $-$ 2451882 & 57.62 $\pm$ 0.21 & 0.005 $\pm$ 0.002 & 148 \\
\enddata
\end{deluxetable}
\clearpage

\begin{deluxetable}{ccl} 
\tablewidth{0pc}
\footnotesize
\tablenum{7}
\tablecaption{Blazhko periods from the literature}
\tablehead{
\colhead{Julian Date} & \colhead{Blazhko Period (days)} & \colhead{Ref.}
          }
\startdata
2432624 $-$ 2434309 & 57.41 & \citet{M53} \\
2417017 $-$ 2434946 & 57.41 & \citet{K58} \\
2438882 $-$ 2441564  & 58.316 & \citet{B73} \\
2441424 $-$ 2442338 & 58.15 & \citet{P75} \\
2417030 $-$ 2438500 & 57.401 $\pm$ 0.016 (max) & \citet{S75} \\
2417030 $-$ 2438500 & 57.396 $\pm$ 0.014 (O-C) & \citet{S75} \\
2438500 $-$ 2441905 & 58.387 $\pm$ 0.076 (max) & \citet{S75} \\
2438500 $-$ 2441905 & 58.33 $\pm$ 0.18 (O-C) & \citet{S75} \\
2443882 $-$ 2442269  & 58.318 & \citet{MT75} \\
\enddata
\end{deluxetable}
\clearpage

\begin{deluxetable}{cl} 
\tablewidth{0pc}
\footnotesize
\tablenum{8}
\tablecaption{Changing Blazhko Periods in RR Lyrae Stars}
\tablehead{
\colhead{Star} & \colhead{$dP_{Blazhko}/dP_{0}$}
          }
\startdata
XZ Cyg & -8$(\pm 1) \times 10^{3}$ \\
XZ Dra & +7.7$(\pm 1.1) \times 10^{4}$ \\
RV UMa & -1 $\times 10^{6}$ \\
RW Dra & -7 $\times 10^{3}$ \\
\enddata
\end{deluxetable}

\end{document}